\documentclass[runningheads]{llncs}

\usepackage[T1]{fontenc}
\usepackage{graphicx}

\usepackage{amsmath,amssymb,amsfonts}
\usepackage{comment}
\usepackage{stmaryrd}
\usepackage{svrsymbols}
\usepackage{xcolor}
\usepackage{algorithmic}
\usepackage{textcomp}
\usepackage[inline]{enumitem}
\usepackage{subcaption}
\usepackage{framed}
\usepackage{csquotes}

\makeatletter
\renewcommand{\paragraph}{\@startsection{paragraph}{5}{0pt}%
	{-1ex\@plus -1ex \@minus -.2ex}%
	{-1.5ex \@plus -.2ex}%
	{\normalfont\normalsize\bfseries}}
\makeatother

\newcommand{\seclabel}[1] {\label{sec:#1}}
\newcommand{\secref}[1] {Section~\ref{sec:#1}}
\newcommand{\thmlabel}[1] {\label{thm:#1}}
\newcommand{\thmref}[1] {Theorem~\ref{thm:#1}}
\newcommand{\lemlabel}[1] {\label{lem:#1}}
\newcommand{\lemref}[1] {Lemma~\ref{lem:#1}}
\newcommand{\proplabel}[1] {\label{prop:#1}}
\newcommand{\propref}[1] {Proposition~\ref{prop:#1}}
\newcommand{\deflabel}[1] {\label{def:#1}}
\newcommand{\defref}[1] {Definition~\ref{def:#1}}

\newcommand{\applabel}[1] {\label{append:#1}}
\newcommand{\appref}[1] {Appendix~\ref{append:#1}}

\makeatletter
\def\moverlay{\mathpalette\mov@rlay}
\def\mov@rlay#1#2{\leavevmode\vtop{%
		\baselineskip\z@skip \lineskiplimit-\maxdimen
		\ialign{\hfil$\m@th#1##$\hfil\cr#2\crcr}}}
\newcommand{\charfusion}[3][\mathord]{
	#1{\ifx#1\mathop\vphantom{#2}\fi
		\mathpalette\mov@rlay{#2\cr#3}
	}
	\ifx#1\mathop\expandafter\displaylimits\fi}
\makeatother

\newcommand{\rmv}[1]{}

\newcommand{\mycases}[3][Case]{\vspace*{0.07in}\noindent {\bf {#1} #2:} #3}

\newcommand{\cupdot}{\charfusion[\mathbin]{\cup}{\cdot}}

\newcommand{\s}[1] {\mathsf{#1}}
\newcommand{\set}[1] {\{#1\}}
\newcommand{\setpred}[2] {\{#1\: |\: #2\}}

\newcommand{\ceil}[1] {\lceil #1 \rceil}
\newcommand{\genquant} {\errorsym}

\newcommand{\ourlogicFull} {Stack-aware Hyper Computation Tree Logic}
\newcommand{\modltl}{{\sc sHCTL*}}
\newcommand{\hypctl}{{\sc HyperCTL*}}

\newcommand{\defeq}{\mathrel{\mathop:}=}					
\newcommand{\coleq}{\mathrel{\mathop:}\mathrel{\mathop:}=}	

\newcommand{\Alpha}{\mathcal{A}}
\newcommand{\natnums}{\mathbb{N}}				  

\newcommand{\ith}[2][i]{#2(#1)}
\newcommand{\suffix}[2][i]{#2[#1:\;]}
\newcommand{\prefix}[2][i]{#2[\;:#1]}
\newcommand{\substr}[3]{#3[#1 : #2]}

\newcommand{\proj}[2][m]{#2|_{#1}}

\DeclareMathOperator{\union}{\cup}				  
\DeclareMathOperator{\disjunion}{\cupdot}		  
\DeclareMathOperator{\orsymb}{\vee}				  
\DeclareMathOperator{\andsymb}{\wedge}			  
\DeclareMathOperator{\impsymb}{\rightarrow}       
\DeclareMathOperator{\until}{\mathsf{U}}		  
\DeclareMathOperator{\release}{\mathsf{R}}		  
\DeclareMathOperator{\satisfy}{\models}			  
\DeclareMathOperator{\notsatisfy}{\not\!\satisfy}	  
\DeclareMathOperator{\existspi}{\exists\pathvar.}	  
\DeclareMathOperator{\forallpi}{\forall\pathvar.}	  
\DeclareMathOperator{\diamondsymb}{\mathsf{F}}	  
\DeclareMathOperator{\squaresymb}{\mathsf{G}}		  

\newcommand{\next}{\mathsf{X}}					  
\newcommand{\apsi}{A\spformula}					  
\newcommand{\epsi}{E\spformula}					  

\newcommand{\ctlformula}{\theta}		
\newcommand{\spformula}{\psi}			
\newcommand{\exec}{\sigma}
\newcommand{\pathvar}{\pi}
\newcommand{\pathvarprime}{\ensuremath{\pi'}}

\newcommand{\alphabet}{\Sigma}   		
\newcommand{\alphele}{a}        	
\newcommand{\word}{w}                	
\newcommand{\emptyword}{\varepsilon}       
\newcommand{\lang}{\mathcal{L}}  		

\newcommand{\PDS}{\mathcal{P}}				 
\newcommand{\PDSexpanded}{(\pdsstates, \stalph, \initpdsstate, \PDSrel, \labelfct)}

\newcommand{\ctrlstates}{Q}                  
\newcommand{\ctrlstate}{q}                   
\newcommand{\initctrlstate}{\ctrlstate_{\s{in}}}  

\newcommand{\pdsstates}{S}
\newcommand{\pdsstate}{s}
\newcommand{\initpdsstate}{\pdsstate_{\s{in}}}

\newcommand{\stalph}{\Gamma}                 
\newcommand{\stword}{\alpha}                 
\newcommand{\stele}{a}
\newcommand{\stbot}{\bot}                 

\newcommand{\call}{\s{call}}
\newcommand{\return}{\s{ret}}
\newcommand{\internal}{\s{int}}

\newcommand{\PDSrel}{\Delta}                 
\newcommand{\PDSrelexp}{\PDSrel_{\internal} \disjunion \PDSrel_{\call} \disjunion \PDSrel_{\return}}

\newcommand{\PDSintexp}{\pdsstates \times \pdsstates}
\newcommand{\PDScallexp}{\pdsstates \times (\pdsstates \times \stalph)}
\newcommand{\PDSreturnexp}{(\pdsstates \times \stalph) \times \pdsstates}

\newcommand{\PDSlabel}[1][(\s{o},\pdsstate,\stele,\pdsstate')]{\xrightarrow{#1}}

\newcommand{\atomprops}{\mathsf{AP}}
\newcommand{\atomprop}{a}
\newcommand{\labelfct}{L}

\newcommand{\stack} {\s{stack}}
\newcommand{\state} {\s{state}}
\newcommand{\topstk} {\s{top}}
\newcommand{\tr}{\s{tr}}
\newcommand{\alt}{{\s{fc}}}
\newcommand{\odd} {\s{odd}}
\newcommand{\even} {\s{even}}

\newcommand{\config}{c}                  
\newcommand{\initconfig}{\config_{\s{in}}}                  
\newcommand{\configPDS}{\s{Conf_{\PDS}}}  

\newcommand{\pathequi}{{equivalent}}

\newcommand{\treq} {\simeq}

\newcommand{\boolsymb}{\mathcal{B}}

\newcommand{\tru}{\s{true}}
\newcommand{\fls} {\s{false}}
\newcommand{\dual}[1] {\s{dual}(#1)}

\newcommand{\transPDS}[1][\PDS]{\llbracket #1 \rrbracket}

\newcommand{\transPDSexpanded}{(\configPDS, \initconfig, \transPDSrel, \atomprops, \labelfct)}
\newcommand{\transPDSrel}{\longrightarrow}    

\newcommand{\parityfct}{\s{parity}}

\newcommand{\auto}{\Alpha}
\newcommand{\autotrans}{\delta}

\newcommand{\spsymbol}{	\dagger}
\newcommand{\pathvars}{\mathcal{V}}  
\newcommand{\pathvardagger}{\mathcal{V}^{\dagger}} 
\newcommand{\pathvarsele}{\pi}
\newcommand{\paths}{\s{Paths}} 
\newcommand{\pathenv}{\Pi} 
\newcommand{\pathenvexp}{\pathvardagger \rightarrow \paths(\transPDS)\union\{\call, \internal, \return\}^\omega}

\newcommand{\stprof}{\rho}                      
\newcommand{\pathprof}{\s{pr}}			
\newcommand{\stprofform}{{\stprofadj} formula}
\newcommand{\stprofadj} {cognate}
\newcommand{\basicfml} {basic}

\newcommand{\vpalphexpanded}{\alphabet_{\call} \disjunion \alphabet_{\internal} \disjunion \alphabet_{\return}}

\newcommand{\acceptstates}{\ctrlstates_{F}}

\newcommand{\autoexpanded}{(\ctrlstates, \initctrlstate, \alphabet, \autotrans, \parityfct)}

\newcommand{\autoVPA}{N}
\newcommand{\autotransVPA}{\delta}
\newcommand{\autoexpandedVPA}{(\ctrlstates, \initctrlstate, \alphabet, \stalph, \stbot, \autotransVPA, \acceptstates)}
\newcommand{\autotransexpandedVPA}{\autotransVPA_{\call} \disjunion \autotransVPA_{\return} \disjunion \autotransVPA_{\internal}}

\newcommand{\nxt} {\rightarrow}
\newcommand{\anxt} {\curvearrowright}

\newcommand{\enc}{\s{enc}}

\newcommand{\inpsymb}{\zeta}		

\newcommand{\ptime} {\textsf{PTIME}}

\newcommand{\dtime}[1] {\textsf{DTIME}(#1)}
\newcommand{\aspace}[1] {\textsf{ASPACE}(#1)}

\newcommand{\rght}{\rightarrow}
\newcommand{\lft}{\leftarrow}
\newcommand{\tmmv}[1][i]{\vdash_{#1}}
\newcommand{\cpy}[2]{[#2]_{#1}}
\newcommand{\sep}{\rhd}

\newcommand{\qptl} {QPTL}

\newcommand{\ctl}{{\sc CTL*}}
\newcommand{\ltl}{{\sc LTL}}
\newcommand{\hypltl}{{\sc HyperLTL}}

\newcommand{\Sta}{Stack-aware}
\newcommand{\sta}{stack-aware}
\newcommand{\rmodltl}{{\sc sHLTL}}

\newcommand*{\short}{}
\newcommand*{\submittedshort}{}

\newcommand{\QPTLchecking}{the QPTL path checking problem }

\begin{document}
	
	\title{Stack-Aware Hyperproperties\thanks{
	{
	Ali Bajwa was partially supported by NSF CNS 1553548. Rohit Chadha was partially supported by NSF CNS 1553548 and NSF SHF 1900924. Mahesh Viswanathan and Minjian Zhang were partially supported by NSF SHF 1901069 and NSF SHF 2007428.}}
	}
	\author{Ali Bajwa\inst{2} \and
		Minjian Zhang\inst{1} \and
		Rohit Chadha\inst{2} \and
		Mahesh Viswanathan\inst{1}}
	
	\authorrunning{A. Bajwa et al.}
	
	\institute{University of Illinois Urbana-Champaign, USA \and
	University of Missouri in Columbia, USA}
	
	\maketitle
	
	\begin{abstract}
	A hyperproperty relates executions of a program and is used to formalize security objectives such as confidentiality, non-interference, privacy, and anonymity. Formally, a hyperproperty is a collection of allowable sets of executions. A program violates a hyperproperty if the set of its executions is not in the collection specified by the hyperproperty. The logic {\hypctl} has been proposed in the literature to formally specify and verify hyperproperties. The problem of checking whether a finite-state program satisfies a {\hypctl} formula is known to be decidable. However, the problem turns out to be undecidable for procedural (recursive) programs. Surprisingly, we show that decidability can be restored if we consider restricted classes of hyperproperties, namely those that relate only those executions of a program which have the same call-stack access pattern. We call such hyperproperties, \emph{stack-aware hyperproperties.} Our decision procedure can be used as a proof method for establishing security objectives such as noninference for recursive programs, and also for refuting security objectives such as observational determinism. Further, if the call stack size is observable to the attacker, the decision procedure provides exact verification.
		
	\end{abstract}
	
	
	\section{Introduction}
    
	
	\ifdefined\short
	Temporal logics {\hypltl} and {\hypctl}~\cite{templogichyperprop-clarkson} were designed to express and reason about security guarantees that are 
	\emph{hyperproperties}~\cite{hyperprop-clarkson}. A hyperproperty~\cite{hyperprop-clarkson} is a security guarantee that does not depend solely on individual
	executions. Instead, a hyperproperty relates multiple executions.  For example, non-interference, a confidentiality property, states that any \emph{two} executions of a program that differ only in high-level security inputs must have the same \emph{low}-security observations.
	As pointed out in~\cite{hyperprop-clarkson}, several security guarantees are hyperproperties.
       The logic  {\hypctl} subsumes {\hypltl}, and 
        the problem of checking a finite-state system against a {\hypctl} formula is decidable~\cite{templogichyperprop-clarkson}. 
        
		\else

The design of computer systems/programs has proved to be error-prone, and  \emph{subtle} flaws are often discovered after these systems have been deployed. Since such flaws may have serious consequences, automated verification techniques that reason about  correctness of such systems have been built. In such approaches, the computer systems, the desired security guarantees and the possible attackers are modeled in some formal logic and then their correctness proved or refuted using logical rules of inference. The logical reasoning could take the form of automated theorem proving, abstract interpretation, type checking or model checking. 

    As pointed out by~\cite{obs-mclean,hyperprop-clarkson}, several security guarantees are not \emph{trace-based}, and in fact, relate multiple traces. For example, non-interference which is a confidentiality property, states that any \emph{two} executions of a program that differ only in high-level security inputs must have the same \emph{low}-security observations. Such security guarantees are \emph{hyperproperties}~\cite{hyperprop-clarkson}, i.e., they are properties of sets of executions instead of a single execution. The temporal logics, {\ltl} and {\ctl}, were extended to {\hypltl} and {\hypctl} respectively in~\cite{templogichyperprop-clarkson} to express hyperproperties. The primary difference is that {\hypltl} and {\hypctl} have explicit path variables, and  are thus able to   quantify over multiple paths simultaneously. As in the case of trace-based properties, {\hypctl} subsumes {\hypltl}. The problem of checking a finite-state system against a {\hypctl} formula is shown to be decidable in~\cite{hyperprop-clarkson}. 
	\fi
 

	 In this paper, we consider the problem of model checking procedural (recursive) programs against security hyperproperties. Recall recursive programs are naturally modeled as a pushdown system. Unlike the case of finite-state transition systems, the problem of checking whether a pushdown system satisfies a {\hypctl} formula is undecidable~\cite{model-pommellet}.
	 %
		In contrast, {\ctl} model checking is decidable for pushdown systems~\cite{pd-bouajjani,pd-walukiewicz}.
	
	\paragraph{Our contributions.}
	We consider restricted classes of hyperproperties for recursive programs, namely those that  relate only those executions that have the same \emph{call-stack access pattern}. Intuitively, two executions have the same stack access pattern if they access the call stack in the same manner at each step, i.e., if in one execution there is a push (pop) at a point, then there is a push (pop) at the same  point in the other execution. Observe that if two executions have the same stack access pattern, then their stack sizes are the same at all times. We call such hyperproperties, \emph{stack-aware hyperproperties.}
	
	In order to specify {\sta} hyperproperties, we extend {\hypctl}  to  {\modltl}. 
    The logic  {\modltl}   has a two level syntax. At the first level, the syntax is
	identical to {\hypctl}  formulas, and is interpreted over executions of the pushdown system with the same stack access pattern. At the top-level, we  quantify over different stack access patterns. The formula $\epsi$ is true if for some stack access pattern $\stprof$ of the system, the pushdown system restricted to executions with stack access pattern $\stprof$ satisfies the {\hypctl} formula $\spformula.$ The formula $\apsi$ is true if for each stack access pattern $\stprof$ of the system, the pushdown system restricted to executions with stack access pattern $\stprof$ satisfies the {\hypctl} formula $\spformula.$ 
	See Figure~\ref{fig:hctlvsshctl2} on Page \pageref{fig:hctlvsshctl2} for a side-by-side comparison of the syntax for {\hypctl}  and {\modltl}. {\hypltl} is extended to {\rmodltl} similarly. Please note that {\modltl} subsumes {\rmodltl}, and that {\modltl} ({\rmodltl}) coincides with {\hypctl} ({\hypltl}) for finite state systems as all executions of finite state systems have the same stack access pattern.  
	
	We show that the  model checking problem for {\modltl} is decidable. We demonstrate three different ways this result can aid in verifying recursive programs. First, for  security guarantees such as noninference~\cite{noninference-mclean}, which are expressible in the $\forall \exists^*$ fragment of {\hypltl}, we can use the model checking algorithm to establish that a recursive program satisfies the {\hypltl} property. 
	Secondly, for the $\forall^*$ fragment of {\hypltl}, the model checking algorithm can be used to detect security flaws by establishing that a recursive program does not satisfy security guarantees. Observational determinism~\cite{obs-mclean,obs-myers} is an example of such a property. Finally, when the attacker can observe stack access patterns (or, equivalently, stack sizes), we can get exact verification for several properties. 
	The assumption of the attacker observing stack access patterns holds, for example, in the program counter security model~\cite{Molnaretal:2006} in which the attacker has access to program counters at each step. As argued in~\cite{Molnaretal:2006}, the program security model is appropriate to capture  control-flow side channels such as those arising from timing behavior and/or disclosure of errors. 
	
	The decision procedure uses an automata-theoretic approach inspired by the model checking algorithm for finite state systems and {\hypctl} given in~\cite{modelchecking-finkbeiner}. Since stack-aware hyperproperties relate only executions with the same stack access-pattern, a set of executions with the same stack access pattern can be encoded as a word over a \emph{pushdown} alphabet,~\footnote{A pushdown alphabet is an alphabet that is partitioned into three sets: a set of call symbols, a set of internal symbols, and a set of return symbols. See \secref{vpa-aja}.} and  the problem of model checking a {\modltl} formula can be reduced to the problem of checking emptiness of a \emph{non-deterministic visibly pushdown automaton (NVPA)} over infinite words~\cite{vpl-alur}. 
	The reduction of the model checking problem to the emptiness problem is based on a compositional construction of an automaton for each sub-formula which accepts exactly the set of assignments to path variables that satisfy the sub-formula. For this construction to be optimal, we carefully leverage two equi-expressive classes of automata on infinite words, namely NVPAs and \emph{1-way alternating jump automata (1-AJA)}~\cite{vpa-bozzelli}. The model checking algorithm for {\modltl} against procedural programs has a complexity that is very close to the complexity of model checking finite state systems against {\hypctl}. If $g(k,n)$ denotes a tower of exponentials of height $k$, where the top most exponent is $\s{poly}(n)$, then for a formula with 
	{formula complexity} $r$,~\footnote{Our definition of 
	{formula complexity} is roughly double the usual notion of quantifier alternation. For a precise definition, see \defref{alt-depth}.} and a system and formula whose size is bounded by $n$, our algorithm is in $\dtime{g(\ceil{\frac{r}{2}},n)}$. In comparison, model checking finite state systems against {\hypctl} is in $\textsf{NSPACE}(g(\ceil{\frac{r}{2}}-1,n))$. This slight difference in complexity is consistent with checking other properties like invariants for finite state systems ($\textsf{NL}$) versus procedural programs ($\textsf{P}$).
	
	We also prove that our model checking algorithm is optimal by proving a matching lower bound. Our proof showing $\dtime{g(\ceil{\frac{r}{2}},n}$-hardness of the model checking problem for formulas with 
	{(formula) complexity} $r$, relies on reducing the membership problem for $g(\ceil{\frac{r}{2}}-1,n)$ space bounded \emph{alternating Turing machines} (ATM) to the model checking problem. The reduction requires identifying an encoding of computations of ATMs, which are trees, as strings that can be guessed and generated by pushdown systems. The pushdown system we construct for the model checking problem guesses potential computations of the ATM, while the {\modltl} formula we construct checks if the guessed computation is a valid accepting computation.
	


	\paragraph{Related work.}
	Clarkson and Schneider introduced \emph{hyperproperties}~\cite{hyperprop-clarkson} and demonstrated their need to capture complex security properties. 
	Temporal logics {\hypltl} and {\hypctl}, that describe hyperproperties, were introduced by Clarkson et al.~\cite{templogichyperprop-clarkson}. They also characterized the complexity of model checking finite state transition systems against {\hypctl} specifications by a reduction to the satisfiability problem of QPTL~\cite{qptl-sistla}. Subsequently, other model checking algorithms for verifying finite state systems against {\hypctl} properties have been proposed~\cite{modelchecking-finkbeiner,mchypext-finkbeiner}. 
	{Tools that check satisfiability~\cite{eahyp-finkbeiner} and runtime verification~\cite{rvhyp-finkbeiner} for {\hypltl} formulas have also been developed.}
	%
	Finkbeiner et al. introduced the automata-theoretic approach to model checking {\hypctl} for finite-state systems~\cite{modelchecking-finkbeiner}. 
	
	The model checking problem for {\hypltl}, and consequently {\hypctl}, was shown to be undecidable for pushdown systems in~\cite{model-pommellet}.
	For restricted fragments of {\hypltl}, Pommellet and Tayssir~\cite{model-pommellet} introduced over-approximations and  under-approximations to 
	establish/refute that a pushdown system satisfies 
	a {\hypltl} formula in those fragments.  Gutsfeld et al. introduced stuttering $H_\mu$, a \emph{linear} time logic for checking asynchronous hyperproperties for recursive programs in~\cite{recursive-gutsfeld}. 
	The authors present complexity results for the model checking problem under an assumption of \emph{fairness}, and a restriction of \emph{well-alignment}. While the restriction to paths with the same \emph{stack access pattern} is similar to the  well-alignment restriction, we do not assume any fairness condition to establish decidability. However, as {\modltl} is a branching time logic and only considers synchronous hyperproperties, the two logics are not directly comparable. It is also worth mentioning that 
	the branching nature of {\modltl} requires us to {\lq\lq copy\rq\rq} a  potentially unbounded
	stack, from the most  recently quantified path variable, when assigning a path to the  {\lq\lq current\rq\rq} quantified path variable. In contrast, all path assignments in~\cite{recursive-gutsfeld} start with an empty stack.
	
	\ifdefined\submittedshort
An extended abstract of this paper appears in the 29th International Conference on Tools and Algorithms for the Construction and Analysis of Systems (TACAS)~\cite{tacaspaper2}.
\else
For lack of space reasons, some proofs are omitted and can be located in~\cite{fullpaper}.
\fi

	
	
	\section{Motivation}
	\label{sec:motivation}

	Clarkson and Schneider~\cite{hyperprop-clarkson}  argue  that many important \emph{security} guarantees are expressible only as \emph{hyperproperties}. 
We discuss two examples of security hyperproperties, and the logics {\hypltl} and {\hypctl} used to specify them.


\sloppy
	\paragraph{Hyperproperties and temporal logics.} 
	
	We discuss two variants of non-interference~\cite{goguen-meseguer} that model confidentiality requirements. 
	In non-interference, the inputs of a system are partitioned into
\emph{low}-level input security variables and \emph{high}-level input security variables. The attacker is assumed to know the values of low-level security inputs. 
During an execution, the attacker can observe  parts of the system configuration such as system outputs, or the memory usage. A system satisfies  \emph{non-interference} if the attacker cannot deduce the values of high-level inputs from the low-level observations. To formally specify the variants, 
we  use the logic {\hypltl}~\cite{templogichyperprop-clarkson},  a fragment of the logic {\hypctl}~\cite{templogichyperprop-clarkson}. The precise syntax of  {\hypltl} and {\hypctl} is shown in Fig.~\ref{fig:hctlvsshctl2}. 
In the syntax,
	$\pi$ is a path variable and the formula $a_\pi$ is true if the proposition $a$ is true along the path \lq\lq $\pi$\rq\rq. Intuitively, the formula  $\existspi \psi$ is existential quantification over paths, and is  true if there is a path that can be assigned to $\pi$ such that $\psi$ is true. Universal quantification ($\forallpi \psi$), and other logical connectives such as conjunction ($\andsymb$), implication ($\impsymb$), equivalence ($\leftrightarrow$) and the temporal operators $\squaresymb$ and $\diamondsymb$ can be defined in the standard way. By having  explicit path variables, {\hypltl} and {\hypctl} allow quantification over multiple paths simultaneously.

\ifdefined\short

	\begin{example}
	\label{exam:ni}
	
The first variant,	noninference~\cite{noninference-mclean}, states that for each  execution $\exec$ of a program, there is another execution $\exec'$ such that \begin{enumerate*}[label=(\alph*)] \item $\exec'$ is obtained from $\exec$ by replacing the high-level security inputs by a dummy input, and \item $\exec$ and $\exec'$ have the same low-level observations. \end{enumerate*} Noninference is a hyperliveness property~\cite{templogichyperprop-clarkson,hyperprop-clarkson}.
	
	Let us assume that the low-level observations of a configuration are 
	determined by the values of the propositions in $L=\set{\ell_1,\cdots \ell_m}.$ 
	As shown in~\cite{templogichyperprop-clarkson}, noninference is expressible  by the {\hypltl} formula:
	 $ \mathsf{NI}\stackrel {\text{def}}{=} \forallpi \exists\pathvar'. (\squaresymb \lambda_{\pi'}) \andsymb  \pi \equiv_L \pi'.$
	 Here $\squaresymb \lambda_{\pi'} $ expresses that \emph{$\squaresymb$lobally} (or in each configuration of the execution) the high input of $\pi'$ is the dummy input $\lambda$, 
	 and  $\pi \equiv_L \pi' \stackrel {\text{def}}{=} \squaresymb (\andsymb_{\ell\in L} (\ell_\pi \leftrightarrow \ell_{\pi'}))$ expresses that $\pi$ and $\pi'$ have the same low-level observations. 
	 \end{example}
	 \begin{example}
	 \label{exam:od}
The second variant, observational determinism~\cite{obs-mclean,obs-myers}, states that any two executions that have the same low-level initial inputs, must have the same low-level output observations. Observational determinism is a hypersafety property~\cite{templogichyperprop-clarkson,hyperprop-clarkson}, and is also expressible in {\hypltl} using the formula~\cite{templogichyperprop-clarkson}:
 	$\mathsf{OD} \stackrel {\mbox{def}}{=}\forallpi \forall\pathvarprime. (\pi[0] \equiv_{L,in} \pi'[0]) \impsymb  \pi \equiv_{L,out} \pi'.$ Here $\equiv_{L,in}$ and $\equiv_{L,out}$ express the fact that $\pi$ and $\pi'$
	have the same low-security inputs and outputs respectively.

	\end{example}
	
	\else
	We illustrate the logic with  two variants of non-interference~\cite{goguen-meseguer} that models confidentiality requirements. We assume that the inputs of a system are partitioned into
\emph{low}-level input security variables and \emph{high}-level input security variables. The values of low-level security input variables are assumed to be known to the attacker. 
As the system executes, some parts of the system configuration such as outputs are observable to the attacker. A system is said to satisfy  \emph{non-interference} if the attacker cannot deduce the values of high-level inputs from the low-level observations.

	\begin{example}
	\label{exam:properties}
	
The first variant,	noninference~\cite{noninference} states that for each  execution $\exec$ of a program, there is another execution $\exec'$ such that \begin{enumerate*}[label=(\alph*)] \item $\exec'$ is obtained from $\exec$ by replacing the high-level security inputs by a dummy input, and \item $\exec$ and $\exec'$ have the same low-level observations. \end{enumerate*} Noninference is a hyperliveness property~\cite{templogichyperprop-clarkson,hyperprop-clarkson}.
	
	Let the low-level observations of a configuration be modeled using propositions $L=\set{\ell_1,\cdots \ell_m}.$ That is, let us assume that the low-level observations of a configuration be completely determined by the truth values of the propositions in the set $L.$ 
	As~\cite{templogichyperprop-clarkson} show, noninference  is expressible in {\hypltl} in using the formula:
	 $ \mathsf{NI}\stackrel {\mbox{def}}{=} \forallpi \existspi' (\squaresymb \lambda_{\pi'}) \andsymb  \pi \equiv_L \pi'.$
	 Here $\squaresymb \lambda_{\pi'} $ expresses that all high inputs in the current configuration are the dummy input $\lambda$ and  $\pi \equiv_L \pi'$ expresses that $\pi$ and $\pi'$ have the same low-level observations. Assuming that the low-level observations are modeled using propositions,  the equivalence $\pi \equiv_L \pi'$ can be expressed as  $\squaresymb (\andsymb_{\ell\in L} (\ell_\pi \Leftrightarrow \ell_{\pi'})).$
	\end{example}
	\begin{example}
	\label{exam:od}
	Observational determinism~\cite{observationaldeterminism} states that any two executions that have the same low-level initial inputs must have the same low-level ouput observations. Observational determinsim is a hypersafety property~\cite{templogichyperprop-clarkson,hyperprop-clarkson}.
	As~\cite{templogichyperprop-clarkson} show, observational determinism is also expressible in {\hypltl} using the formula:
	$\mathsf{OD} \stackrel {\mbox{def}}{=}\forallpi \forallpi' (\pi[0] \equiv_{L,in} \pi'[0]) \impsymb  \pi \equiv_{L,out} \pi'.$ Here $\equiv_{L,in}$ and $\equiv_{L,out}$ express the fact that $\pi$ and $\pi'$
	have the same low-security inputs and outputs respectively.   
	\end{example}
\fi

	\paragraph{Procedural (recursive) programs
 and {\Sta} hyperproperties.}
Pushdown systems model procedural  programs that do not dynamically allocate memory, and whose program variables take values in finite domains.
        Unlike finite-state transition systems, the problem of checking whether a  pushdown system satisfies a {\hypctl} formula  is undecidable~\cite{model-pommellet}. 
         However, we identify a natural class of hyperproperties for which the model checking problem becomes decidable. As we shall shortly see, this class of hyperproperties not only enjoys decidability, but is also useful in reasoning about security hyperproperies such as noninference and observational determinism. 
         
	
	We consider a restricted class of hyperproperties for recursive programs, which relate only executions that access the call stack in the same manner, i.e., push or pop at the same time.
	An execution of a pushdown system $\PDS$ is a sequence of configurations (control state + stack) $\exec=\config_1\config_2 \cdots,$ such that 
	the stacks of consecutive configurations $\config_i$ and $\config_{i+1}$ differ only due to the possible presence of an additional element at the top of the stack of either $\config_i$ or $\config_{i+1}$. 
	For such a sequence, we can associate a sequence $\pathprof(\exec)=\s{o}_1 \s{o}_2 \cdots $ such that $\s{o}_i\in \set{\call,\internal,\return}$ such that $\s{o}_i=\call$ ($\return$ respectively) if and only if the stack in $\config_{i+1}$ has one more (less respectively) element than $\config_i.$  The sequence $\pathprof(\exec)$ is said to be the \emph{stack access pattern} of $\exec$. Observe that the stack sizes of two executions with the same stack access pattern evolve in a similar fashion. Thus, equivalently, we can consider this  class of hyperproperties to be the hyperproperties that relate executions with identical memory usage.
	To specify these hyperproperties, we propose the logic {\modltl}  which extends  {\hypctl}. 	
	{\modltl} has a two level syntax. At the innermost level, the syntax is
	identical to that of {\hypctl}  formulas, but is interpreted over executions of the pushdown system with the same stack access pattern. At the outer level, we  quantify over different stack access patterns. Intuitively, the formula $\epsi$ is true if there is a stack access pattern $\stprof$ exhibited by the system such that the set of executions with access pattern $\stprof$ satisfy the hyperproperty $\spformula.$
	The dual formula $\apsi$, defined as $\lnot E \lnot \psi$, is true if for each stack access pattern $\stprof$ exhibited by the system, the set of all executions with stack access pattern $\stprof$  satisfy $\spformula.$ The syntax of {\rmodltl} is obtained from {\hypltl} in a similar fashion. Please see Fig.~\ref{fig:hctlvsshctl2} on Page~\pageref{fig:hctlvsshctl2} for
	a side-by-side comparison of the syntax of {\hypctl} ({\hypltl}) and {\modltl} ({\rmodltl}).
	Unlike {\hypctl}, we show  that the problem of checking {\modltl} is decidable for pushdown systems (\thmref{main}).
	Formal definitions of stack access patterns, syntax and semantics of {\modltl} are in \secref{shctl}. 
	
	For the rest of the paper, hyperproperties expressible in {\modltl} will be called \emph{stack-aware hyperproperties}. Restricting to stack-aware hyperproperties is useful in verifying security guarantees of recursive programs as discussed below.

	\paragraph{Proving $\forall\exists^*$ hyperproperties.}
	The  \emph{noninference} property (Example~\ref{exam:ni}) can be expressed in {\hypltl} as $ \mathsf{NI}\stackrel {\mbox{def}}{=} \forallpi \existspi' (\squaresymb \lambda_{\pi'}) \andsymb  \pi \equiv_L \pi'$.  Consider the {\rmodltl} formula $A (\mathsf{NI})$ obtained by putting an $A$ in front $\mathsf{NI}$. A pushdown system satisfies $A (\mathsf{NI})$ only if for each execution $\exec$ of the system, there is another execution $\exec'$ with \emph{the same stack access pattern as $\exec$} such that  $\exec,\exec'$ together satisfy $(\squaresymb \lambda_{\exec'}) \andsymb  \exec \equiv_L \exec'.$ Thus, if the pushdown system satisfies the  {\rmodltl} formula $A (\mathsf{NI})$, then it also satisfies  noninference.  Thus, a decision procedure for {\rmodltl} can be used to prove that a recursive program satisfies noninference. 
	
	\sloppy
	The above observation generalizes to {\hypltl} formulas of the form $\spformula=$ $\forall \pi. \exists \pi_1. \ldots \exists \pi_k. \spformula'$ --- if a system satisfies the {\rmodltl} formula $\apsi$ then it must also satisfy the {\hypltl} formula $\spformula.$  Though the model checking problem is undecidable for pushdown systems even when restricted to such {\hypltl} formulas, we gain decidability by restricting the search space for $\pi, \pi_1,\ldots, \pi_k$.
	
	

	\paragraph{Refuting $\forall^*$ hyperproperties.}
		\emph{Observational determinism} (Example~\ref{exam:od}) can be written in {\hypltl} as
 	$\mathsf{OD} \stackrel {\mbox{def}}{=}\forallpi \forall\pathvarprime. (\pi[0] \equiv_{L,in} \pi'[0]) \impsymb  \pi \equiv_{L,out} \pi'.$ Consider the {\rmodltl} formula $A (\mathsf{OD})$.
 	A pushdown system  \emph{fails} to satisfy the
     {\rmodltl} formula $A (\mathsf{OD})$ only if there is a stack access pattern $\stprof$ and executions $\exec_1$ and $\exec_2$ with stack access pattern $\stprof$ such that the pushdown system does not satisfy $(\exec[0] \equiv_{L,in} \exec'[0]) \impsymb  \exec \equiv_{L,out} \exec'.$
 	
 	This observation generalizes to {\hypltl} formulas of the form $\spformula=\forall \pi_1. \ldots \forall \pi_k. \spformula'$ --- if a pushdown system fails to satisfy the {\rmodltl} formula $\apsi$ then it does not satisfy $\spformula.$ Even though model checking pushdown systems against such restricted specifications is undecidable, our decision procedure can be used to show that a recursive program does not meet such properties.
 	%

	
	
	\paragraph{Exact verification when stack access pattern is observable.} 
	Often, it is reasonable to assume that the attacker can observe the stack access pattern. For example, in the program counter security model~\cite{Molnaretal:2006}, the attacker has access to
	the program counter transcript, i.e.,  the sequence of program counters during an execution. Access to the program counter transcript implies that the attacker can observe stack access pattern. The assumption that the program counter transcript is observable
	helps model control flow side channel attacks which include timing attacks and error disclosure attacks ~\cite{Molnaretal:2006}.
	{\modltl} can be used to verify security guarantees in this security model. For example,
	the
   {\modltl} formula $ A(\,\mathsf{NI})$ models noninference faithfully by introducing a unique proposition for each control state. Observational determinism can also be verified in this model by suitably transforming the pushdown automaton. 
		
Another scenario in which stack access patterns are observable is when the attacker can observe the memory usage of a program in terms of stack size. As observing stack size may lead to information leakage, stack size should be considered a low-level observation.
Since the stack size can be unbounded, it
cannot be modeled as a proposition. {\modltl}, however, can still be used to verify security guarantees in this case. For  example, $A(\,\mathsf{NI})=A( \forallpi \existspi' (\squaresymb \lambda_{\pi'}) \andsymb  \pi \equiv_L \pi')$ faithfully models non-inference as semantics of {\modltl} forces $\pi$ and $\pi'$ to have the same call-stack size in addition to other low-level observations. Once again, observational determinism can also be verified in this model by suitably transforming the pushdown automaton.

	\ifdefined\short
	\section{Stack-aware Hyper Computation Tree Logic ({\modltl})}
	\label{sec:shctl}
	
	Stack-aware Hyper Computation Tree Logic ({\modltl}), and its sub-logic Stack-aware Hyper Linear Temporal Logic ({\rmodltl}) are formally presented. We begin by establishing some conventions over strings.

	\paragraph{Strings.}
	A \emph{string/word} $\word$ over a finite alphabet $\alphabet$ is a sequence $w = \alphele_0\alphele_1\cdots$ of finite or infinitely many symbols from $\alphabet$, i.e., $\alphele_i \in \alphabet$ for all $i$. The \emph{length} of a string $w$, denoted $|\word|$, is the number of symbols appearing in it --- if $\word = \alphele_0\alphele_1\cdots\alphele_{n-1}$ is finite then $|\word| = n$, and if $\word = \alphele_0\alphele_1\cdots$ is infinite then $|\word| = \omega$. The \emph{unique} string of length $0$, the \emph{empty string}, is denoted $\emptyword$. For a string $\word = \alphele_0\alphele_1\cdots \alphele_i\cdots$, $\ith{\word} = \alphele_i$ denotes the $i$th symbol, $\prefix{\word} = \alphele_0\alphele_1\cdots\alphele_{i-1}$ denotes the prefix of length $i$, $\suffix{\word} = \alphele_i\alphele_{i+1}\cdots$ denotes the suffix of $\word$ starting at position $i$, and $\substr{i}{j}{\word} = \alphele_i\alphele_{i+1}\cdots\alphele_{j-1}$ denotes the substring from position $i$ (included) to position $j$ (not included). Thus $\suffix[0]{\word} = \word$. By convention, when $i \leq 0$, we take $\prefix{\word} = \emptyword$. Over $\alphabet$, the set of all finite strings is denoted $\alphabet^*$, and the set of all infinite strings is denoted $\alphabet^\omega$. For a finite string $u$ and a (finite or infinite) string $v$, $uv$ denotes the \emph{concatenation} of $u$ and $v$.

	 \subsection{Pushdown Systems}
	\rmv{ 
	  As described in Section~\ref{sec:motivation}, 	
	the semantics of a sequential, recursive program is most naturally modeled using pushdown systems, which have finitely many control states and an unbounded stack.
	We will consider pushdown systems where the control states are labeled by facts or propositions in our logic that are true in a particular control state. Formally, an \emph{$\atomprops$-labeled pushdown system} is a tuple $\PDS = \PDSexpanded$, where $\pdsstates$ is a (finite) set of \emph{control states}, $\stalph$ is a finite set of \emph{stack symbols}, $\initpdsstate \in \pdsstates$ is the \emph{initial control state}, $\labelfct: \pdsstates \to 2^{\atomprops}$ is the \emph{labeling function}, and $\PDSrel$ is the transition relation. The transition relation $\PDSrel = \PDSrelexp$ is the disjoint union of three transition relations --- \emph{internal transitions} $\PDSrel_{\internal} \subseteq \PDSintexp$ where the stack is not changed, \emph{call transitions} $\PDSrel_{\call} \subseteq \PDScallexp$ where a single symbol is \emph{pushed} onto the stack, and \emph{return transitions} $\PDSrel_{\return} \subseteq \PDSreturnexp$ where a single symbol is \emph{popped} from the stack. When the label set $\atomprops$ is clear from the context, we simply refer to them as pushdown systems.
	}
	Pushdown systems naturally model for sequential recursive programs. 
	Formally, an \emph{$\atomprops$-labeled pushdown system} is a tuple $\PDS = \PDSexpanded$, where $\pdsstates$ is a finite set of \emph{control states}, $\stalph$ is a finite set of \emph{stack symbols}, $\initpdsstate \in \pdsstates$ is the \emph{initial control state}, $\labelfct: \pdsstates \to 2^{\atomprops}$ is the \emph{labeling function}, and $\PDSrel$ is the transition relation. The transition relation $\PDSrel = \PDSrelexp$ is the disjoint union of \emph{internal transitions} $\PDSrel_{\internal} \subseteq \PDSintexp$ where the stack is unchanged, \emph{call transitions} $\PDSrel_{\call} \subseteq \PDScallexp$ where a single symbol is \emph{pushed} onto the stack, and \emph{return transitions} $\PDSrel_{\return} \subseteq \PDSreturnexp$ where a single symbol is \emph{popped} from the stack. When 
	$\atomprops$ is clear from the context, we simply refer to them as pushdown systems.
		\ifdefined\short
		\else
	\begin{remark}
		\textcolor{red}{MV: Remove this for TACAS?} Our definition of pushdown systems, though standard, differs slightly from the presentation of a pushdown automaton found in textbooks. It is tailored to capture programs more naturally. Transitions are not labeled explicitly with inputs because we are not interested in automata languages.
 Second, transitions either push one symbol or pop one symbol or leave the stack unchanged, which is consistent with the way a call stack changes on a call, return, and other steps. 
	\end{remark}
       \fi
			
\noindent{\textbf{Transition System Semantics.}}	We recall the standard semantics of a pushdown system as a transition system. Let us fix a pushdown system $\PDS = \PDSexpanded$. A \emph{configuration} $\config$ of $\PDS$ is a pair $(\pdsstate,\stword)$ where $\pdsstate \in \pdsstates$ and $\stword \in \stalph^*$. The set of all configurations of $\PDS$ will be denoted $\configPDS = \pdsstates \times \stalph^*$. The \emph{labeled transition system} associated with $\PDS$ is $\transPDS \defeq \transPDSexpanded$ where $\initconfig = (\initpdsstate,\emptyword)$ is the \emph{initial configuration}, 
{$\transPDSrel \subseteq \configPDS \times (\set{\call, \return, \internal} \times \pdsstates \times (\stalph \union \set{\emptyword})\times \pdsstates) \times \configPDS$} is the \emph{transition relation}, and $\labelfct$ is the \emph{labeling function} that extends the labeling function of $\PDS$ to configurations as follows: $\labelfct(\pdsstate,\stword) = \labelfct(\pdsstate)$. The transition relation $\transPDSrel$ is defined to capture the informal semantics of internal, call, and return transitions --- for any $ \stword \in \stalph^*$, ($\internal$) $(\pdsstate, \stword) \PDSlabel[(\internal,\pdsstate,\emptyword,\pdsstate')] (\pdsstate', \stword)$ iff $(\pdsstate,\pdsstate') \in \PDSrel_{\internal}$; ($\call$) $(\pdsstate, \stword) \PDSlabel[(\call,\pdsstate,\stele,\pdsstate')] (\pdsstate', \stele\stword)$ iff $(\pdsstate, (\pdsstate', \stele)) \in \PDSrel_{\call}$; and ($\return$) $(\pdsstate, \stele\stword) \PDSlabel[(\return,\pdsstate,\stele,\pdsstate')] (\pdsstate', \stword)$ iff $((\pdsstate, \stele), \pdsstate') \in \PDSrel_{\return}$.
	\rmv{
	The semantics of a pushdown system is described in terms of a labeled transition system with infinitely many states. Let us fix a pushdown system $\PDS = \PDSexpanded$. A \emph{configuration} $\config$ of $\PDS$ is a pair $(\pdsstate,\stword)$ where $\pdsstate \in \pdsstates$ and $\stword \in \stalph^*$. The set of all configurations of $\PDS$ will be denoted $\configPDS = \pdsstates \times \stalph^*$. The \emph{labeled transition system} associated with $\PDS$ is $\transPDS \defeq \transPDSexpanded$ where $\initconfig = (\initpdsstate,\emptyword)$ is the \emph{initial configuration}, $\transPDSrel \subseteq \configPDS \times \PDSrel \times \configPDS$ is the \emph{transition relation}, and $\labelfct$ is the \emph{labeling function} that extends the labeling function of $\PDS$ to configurations as follows: $\labelfct(\pdsstate,\stword) = \labelfct(\pdsstate)$. The transition relation $\transPDSrel$ is defined to capture the informal semantics of internal, call, and return transitions --- for any $ \stword \in \stalph^*$, ($\internal$) $(\pdsstate, \stword) \PDSlabel[(\internal,\pdsstate,\emptyword,\pdsstate')] (\pdsstate', \stword)$ iff $(\pdsstate,\pdsstate') \in \PDSrel_{\internal}$; ($\call$) $(\pdsstate, \stword) \PDSlabel[(\call,\pdsstate,\stele,\pdsstate')] (\pdsstate', \stele\stword)$ iff $(\pdsstate, (\pdsstate', \stele)) \in \PDSrel_{\call}$; and ($\return$) $(\pdsstate, \stele\stword) \PDSlabel[(\return,\pdsstate,\stele,\pdsstate')] (\pdsstate', \stword)$ iff $((\pdsstate, \stele), \pdsstate') \in \PDSrel_{\return}$.
	}
	
	A \emph{path} of $\transPDS$ is an infinite sequence of configurations $\exec = \config_0,\config_1,\ldots$ such that for each $i$, $c_i \PDSlabel c_{i+1}$ for some $\s{o} \in \set{\internal,\call,\return}$, $\pdsstate,\pdsstate' \in \pdsstates$ and $\stele \in \stalph \cup \set{\emptyword}$. The path $\exec$ is said to \emph{start} in configuration $\config_0$ (the first configuration in the sequence). We will use $\paths(\transPDS, \config)$ to denote the set of paths of $\transPDS$ starting in the configuration $\config$ and $\paths(\transPDS)$ to denote all paths of $\transPDS$.
	
	We conclude this section by introducing some notation on configurations. For $\config = (\pdsstate,\stword)$, its \emph{stack height} is $|\stword|$, its \emph{control state} is $\state(\config) = \pdsstate$, and its \emph{top of stack symbol} is $\topstk(\config) = \stele \in \stalph$ if $\stword = \stele\stword'$ and is undefined if $\stword = \emptyword$.

	\subsection{Syntax of {\modltl}}
	
			\begin{figure}[t]
			\centering
				\begin{framed}
			
				$\atomprop \in \atomprops,\pathvar\in\pathvars$\\
				\vspace*{-0.2in}
			\begin{subfigure}[b]{0.40\textwidth}
				\[
				\begin{array}{ccccccccc}
					\spformula & \coleq & \atomprop_\pathvar &\vert& \lnot\spformula &\vert& \spformula \orsymb \spformula &\vert& \next \spformula \\
					 & & &  \vert& \spformula \until \spformula &\vert& \existspi \spformula \\
				\end{array}
				\]
				\vspace*{-0.1in}
				\subcaption{{\hypctl}}
				\label{sub:hctl2}
			\end{subfigure}
		\hfill
			\begin{subfigure}[b]{0.54\textwidth}
				\[
				\begin{array}{ccccccccccccc}
				    \\
					\ctlformula & \coleq &  \epsi &\vert& \lnot \ctlformula &\vert&\ctlformula \orsymb \ctlformula \\ 
					
					\spformula & \coleq & \atomprop_\pathvar &\vert& \lnot\spformula &\vert& \spformula \orsymb \spformula &\vert& 
					\next \spformula &\vert& \spformula \until \spformula &\vert& \existspi \spformula
				\end{array}
				\]
				\vspace*{-0.1in}
				\subcaption{{\modltl}}
				\label{sub:shctl2}
			\end{subfigure}
			\end{framed}
		\caption{\footnotesize{BNF for {\hypctl} and {\modltl}. 
		Let  $\forall$ denote $\lnot \exists \lnot$ and $A$ denote $\lnot E\lnot \psi.$ 
		{\hypltl} is the set of {\hypctl} formulas $Q_1 \pi_1. \cdots Q_r \pi_r. \psi$ where $Q_i\in \set{\exists,\forall}$ and $\psi$ is quantifier-free.  {\rmodltl} is the set of {\modltl} formulas $\genquant \varphi$, where $\genquant\in \set{A,E}$ and $\varphi$ is in {\hypltl}.}}
		\label{fig:hctlvsshctl2}
	
	\end{figure}

	Let us fix a  set of atomic propositions $\atomprops$, and a set of path variables, $\pathvars$. The BNF grammar for {\modltl} formulas is given in Figure~\ref{fig:hctlvsshctl2}(b). In the BNF grammar,
	$\atomprop \in \atomprops$ is an \emph{atomic proposition}, $\pathvar$ is a \emph{path variable}, $\spformula$ is a \emph{\stprofform}, and $\ctlformula$ is a {\modltl} formula. 
	The syntax has two levels, with the inner level identical to {\hypctl} formulas, while the outer level allows quantification over different stack access patterns (see \secref{semantics}).
	Also,
	following~\cite{templogichyperprop-clarkson,modelchecking-finkbeiner}, we assume that the until operator $\until$ only occurs within the scope of a path quantifier. 
			

	\begin{remark}
	    We have chosen to not have $A$, the dual of $E$, and conjunction as  explicit logical operators to keep our exposition simple. This choice does makes the automata constructions presented here less efficient for formulas involving conjunction. Adding them explicitly does not pose a technical challenge to our setup and our automata constructions can be extended to handle them explicitly.
	    In addition, we will sometimes use other quantifiers and logical operators to write formulas. Some standard examples include: 
	    $\ctlformula_1 \andsymb \ctlformula_2 = \lnot (\lnot \ctlformula_1 \orsymb \lnot \ctlformula_2)$, where $\ctlformula_i$ ($i \in \set{1,2}$) is either a {\modltl} or {\stprofadj} formula; $\forall\pathvar.\: \spformula = \lnot \existspi \lnot \spformula$; $\diamondsymb \spformula = \tru\until\spformula$, where $\tru = \atomprop_\pathvar \orsymb \lnot\atomprop_\pathvar$; $\squaresymb\spformula = \lnot\diamondsymb\lnot\spformula$.
	\end{remark}

	We call formulas of the form $\genquant\spformula$ (where $\genquant \in \set{A,E}$ and $\spformula$ is a {\stprofform}) \emph{{\basicfml} formulas}. Observe that any {\modltl} formula is a Boolean combination of {\basicfml} formulas. A {\modltl} formula $\ctlformula$ is a \emph{sentence} if in each {\basicfml} sub-formula $\genquant\spformula$, $\spformula$ is a sentence, i.e., every path variable appearing in $\spformula$ is quantified. Without loss of generality, we assume that in any {\stprofform} $\spformula$, all bound  variables in $\spformula$ are renamed to ensure that any path variable is quantified at most once. We will only consider {\modltl} sentences in this paper. The logic {\rmodltl} is the sub-logic of {\modltl}
	consisting of all formulas of the form $\genquant Q_1 \pi_1. \cdots Q_r \pi_r. \psi$
	where $\genquant\in \set{A,E}$, $Q_i\in \set{\exists,\forall}$ and $\psi$ is quantifier free.

	\subsection{Semantics of \modltl}
	\seclabel{semantics}
	
	
	The syntax of {\stprofform}s is identical to that {\hypctl} formulas. Their semantics will be described in a similar manner, in a context where free path variables in the formula are interpreted as executions of a system. However, we will require that the interpretations of every path variable share a \emph{common} stack access pattern --- hence the term \emph{\stprofadj}. Thus, before defining the semantics, we will define what we mean by the \emph{stack access pattern} of a path and a \emph{path environment} that assigns an interpretation to path variables. 
	
	For the rest of this section let us fix a pushdown system $\PDS = \PDSexpanded$. A string $\word \in \set{\call,\internal,\return}^*$ is said to be \emph{well matched} if either $\word = \emptyword$ or $\word = \internal$ or $\word = \call\: u\: \return$ or $\word = uv$, where $u, v \in \set{\call,\internal,\return}^*$ are (recursively) well matched. In a string $\stprof \in \set{\call,\internal,\return}^\omega$, $\ith{\stprof}$ is an \emph{unmatched return}, if $\prefix[i+1]{\stprof} = \word\:\return$, where $\word$ is well matched. We are now ready to present the definition of a stack access pattern.
	
	\begin{definition}[Stack access pattern]
		\sloppy A string $\stprof \in \set{\call,\internal,\return}^\omega$ is a \emph{stack access pattern } if the set $\setpred{i \in \natnums}{\ith{\stprof} \mbox{ is an unmatched return}}$ is finite.
		
		A path $\exec = \config_0 \config_1 \config_2 \cdots \in \paths(\transPDS)$ is said to have a stack access pattern $\stprof = o_0 o_1 \cdots$ (denoted $\pathprof(\exec) = \stprof$) if for every $i$:
		\begin{enumerate*}[nosep,label=(\alph*)]
			\item $o_i = \call$ if and only if $\stack(c_{i+1})$ = $\topstk(c_{i+1})\, \stack(c_i)$, 
			\item $o_i = \internal$ if and only if $\stack(c_{i+1}) = \stack (c_i)$, and
			\item $o_i = \return$ if and only if $\stack(c_i) =\topstk(c_i)\, \stack (c_{i+1})$.
		\end{enumerate*}
	\end{definition}
	
	We now present the definition of \emph{path environment} that interprets the free path variables in a {\stprofform} as paths of $\transPDS$ such that they share a common stack access pattern. This plays a key role in defining the semantics of {\modltl}. For a set of path variables $\pathvars$, let $\pathvardagger$ be defined as the set $\pathvars \disjunion \set{\spsymbol}$.
	
	\begin{definition}[Path Environment]
		A \emph{path environment} for pushdown system $\PDS$ over variables $\pathvars$ is function $\pathenv: \pathenvexp$ such that $\pathenv(\spsymbol)$ is a stack access pattern , and for every $\pathvarsele \in \pathvars$, $\pathenv(\pathvarsele) \in \paths(\transPDS)$ with $\pathprof(\pathenv(\pathvarsele)) = \pathenv(\spsymbol)$. When the pushdown system is clear from the context, we will simply refer to it as a path environment over $\pathvars$.
		
		When $\pathvars = \emptyset$, we additionally require that there is a path $\exec \in \paths(\transPDS,\initconfig)$ (where $\initconfig$ is the initial configuration of $\transPDS$) such that $\pathprof(\exec) = \pathenv(\spsymbol)$.
	\end{definition}
	
	
	We introduce some notation related to path environments. 
	Let us fix a path environment $\pathenv$ over variables $\pathvars$. Given a path $\exec \in \paths(\transPDS)$, $\pathenv[\pathvar \mapsto \exec]$ denotes the path environment over $\pathvars \union \set{\pathvar}$ such that $\pathenv[\pathvar \mapsto \exec](\pathvar) = \exec$, and $\pathenv[\pathvar \mapsto \exec](\pathvar') = \pathenv(\pathvar')$, for any $\pathvar' \in \pathvardagger$ with $\pathvar' \neq \pathvar$. Finally, for $i \in \natnums$, $\suffix{\pathenv}$ denotes the \emph{suffix} path environment, where every variable is mapped to the suffix of the path starting at position $i$. More formally, for every $\pathvar' \in \pathvardagger$, $\suffix{\pathenv}(\pathvar') = \suffix{\pathenv(\pathvar')}$.
	
	We now define when a pushdown system $\PDS$ satisfies a {\modltl} sentence $\ctlformula$, denoted $\PDS \satisfy \ctlformula$. The definition of satisfaction of $\ctlformula$ relies on a definition of satisfaction for {\stprofform}s. To inductively to define the semantics of {\stprofform}s, we will interpret free path variables using a path environment. Finally, as in {\hypctl}, it is important to track the most recently quantified path variable because that influences the semantics of $\exists \pi (\cdot)$. Thus satisfaction of {\stprofform}s takes the form $\PDS,\pathenv,\pathvarsele' \satisfy \spformula$, where $\pathvarsele'$ is the most recently quantified path variable, and $\pathenv$ is a path environment over the free variables of $\spformula$. Finally, by convention, we will take $\paths(\transPDS,\pathenv(\spsymbol)(0))$ to mean $\paths(\transPDS,\initconfig)$, where $\initconfig$ is the initial configuration of $\transPDS$~\footnote{The convention is needed because $\pathenv(\spsymbol)(0)$ is not a configuration but an element of the set $\set{\call,\internal,\return}$.}. Below, $\ctlformula, \ctlformula_1,$ and $\ctlformula_2$ are {\modltl} sentences, while $\spformula,\spformula_1,\spformula_2$ are {\stprofform}s.
	\[
		\begin{array}{l}
		    \PDS \satisfy \lnot \ctlformula \text{ iff } \PDS \notsatisfy \ctlformula\\
		    \PDS \satisfy \ctlformula_1 \orsymb \ctlformula_2 \text{ iff } \PDS \satisfy \ctlformula_1 \text{ or } \PDS \satisfy \ctlformula_2 \\

			\PDS \satisfy \epsi \text{ iff } \text{for some path environment } \pathenv \text{ over } \emptyset, \PDS, \pathenv, \spsymbol \satisfy \spformula \\
			
			
			\PDS, \pathenv, \pathvarsele'  \satisfy \atomprop_\pathvar \text{ iff } \atomprop \in \labelfct(\ith[0]{\pathenv(\pathvar)}) \\
			
			\PDS, \pathenv, \pathvarsele' \satisfy \lnot\spformula \text{ iff } \PDS, \pathenv, \pathvarsele' \notsatisfy \spformula \\
			
			\PDS, \pathenv, \pathvarsele' \satisfy \spformula_1 \orsymb \spformula_2 \text{ iff } \PDS, \pathenv, \pathvarsele' \satisfy \spformula_1 \text{ or } \PDS, \pathenv, \pathvarsele' \satisfy \spformula_2 \\
			
			
			\PDS, \pathenv, \pathvarsele' \satisfy \next \spformula \text{ iff } \PDS, \suffix[1]{\pathenv},\pathvarsele' \satisfy \spformula \\
			
			\PDS, \pathenv, \pathvarsele' \satisfy \spformula_1 \until \spformula_2 \text{ iff } \exists i \geq 0 : \PDS,\suffix{\pathenv}, \pathvarsele' \satisfy \spformula_2 \text{ and } \forall j, 0 \leq j < i, \\ 
			\hspace*{3.4cm}\PDS,\suffix[j]{\pathenv}, \pathvarsele' \satisfy \spformula_1 \\
			
			\PDS, \pathenv, \pathvarsele' \satisfy \existspi \spformula \text{ iff } \exists \exec \in \paths(\transPDS, \ith[0]{\pathenv(\pathvarsele')}) \text{ with } \pathprof(\exec)=\pathenv(\spsymbol), \\
			\hspace*{3cm}\text{such that } \PDS,\pathenv[\pathvar \mapsto \exec], \pathvar \satisfy \spformula
		\end{array}
	\]

	\else

	\section{Pushdown Systems}
    As described in Section~\ref{sec:motivation}, 	
	the semantics of a sequential, recursive program is most naturally modeled using pushdown systems, where the stack is used to model the call stack of a program. 
	We introduce this model of pushdown systems formally and establish basic definitions and notations, that will be used throughout the paper.

	\paragraph{Strings.}
	A \emph{string/sequence} $\word$ over a finite alphabet $\alphabet$ is a sequence $w = \alphele_0\alphele_1\cdots\alphele_n\cdots$ of finitely or infinitely many symbols from $\alphabet$, i.e., $\alphele_i \in \alphabet$ for all $i \in \natnums$. The \emph{length} of a string $w$, denoted $|\word|$ is the number of symbols appearing in it --- if $\word = \alphele_0\alphele_1\cdots\alphele_{n-1}$ is finite then $|\word| = n$, and if $\word = \alphele_0\alphele_1\cdots$ is infinite then $|\word| = \omega$. The \emph{unique} string of length $0$, called the \emph{empty string}, will be denoted as $\emptyword$. For a string $\word = \alphele_0\alphele_1\cdots \alphele_i\cdots$, $\ith{\word} = \alphele_i$ denotes the $i$th symbol, $\prefix{\word} = \alphele_0\alphele_1\cdots\alphele_{i-1}$ is the prefix of length $i$, $\suffix{\word} = \alphele_i\alphele_{i+1}\cdots$ denotes the suffix of $\word$ starting at position $i$, and $\substr{i}{j}{\word} = \alphele_i\alphele_{i+1}\cdots\alphele_{j-1}$ denotes the substring from position $i$ (included) to position $j$ (not included). Thus $\suffix[0]{\word} = \word$. By convention, when $i \leq 0$, we take $\prefix{\word} = \emptyword$. Over $\alphabet$, the set of all finite strings is denoted $\alphabet^*$ and the set of all infinite strings by $\alphabet^\omega$. For a finite string $u$ and a (finite or infinite) string $v$, $uv$ denotes the \emph{concatenation} of $u$ and $v$.
	
	\paragraph{Pushdown Systems.}
	A pushdown system has finitely many control states and an unbounded stack. We will consider pushdown systems where the control states are labeled. These labels will correspond to facts or propositions in our logic that are true in a particular control state. Formally, an \emph{$\atomprops$-labeled pushdown system} is a tuple $\PDS = \PDSexpanded$ where $\pdsstates$ is a (finite) set of \emph{control states}, $\stalph$ is a finite set of symbols constituting the \emph{stack alphabet}, $\initpdsstate \in \pdsstates$ is the \emph{initial control state}, $\labelfct: \pdsstates \to 2^{\atomprops}$ is the \emph{labeling function}, and $\PDSrel$ is the transition relation. The transition relation $\PDSrel = \PDSrelexp$ is the disjoint union of three transition relations --- \emph{internal transitions} $\PDSrel_{\internal} \subseteq \PDSintexp$ where the stack is not changed, \emph{call transitions} $\PDSrel_{\call} \subseteq \PDScallexp$ where a single symbol is \emph{pushed} onto the stack, and \emph{return transitions} $\PDSrel_{\return} \subseteq \PDSreturnexp$ where a single symbol is \emph{popped} from the stack. When the label set $\atomprops$ is clear from the context, we will just refer to them simply as pushdown systems.
	
	\begin{remark}
		Our definition of pushdown systems, though standard, differs slightly from the presentation of a pushdown automaton found in textbooks. It is tailored to capture programs more naturally. Transitions are not labeled with inputs because this is used to model a (closed) program and not an algorithm that processes inputs. Second, transitions either push one symbol or pop one symbol or leave the stack unchanged, which is consistent with the way a call stack changes on a call, return, and other steps. 
	\end{remark}
	
	\paragraph{Transition System Semantics.}
	The semantics of a pushdown system is described in terms of a labeled transition system with infinitely many states. Let us fix a pushdown system $\PDS = \PDSexpanded$. A \emph{configuration} $\config$ of $\PDS$ is a pair $(\pdsstate,\stword)$ where $\pdsstate \in \pdsstates$ and $\stword \in \stalph^*$. The set of all configurations of $\PDS$ will be denoted $\configPDS = \pdsstates \times \stalph^*$. The \emph{labeled transition system} associated with $\PDS$ is $\transPDS \defeq \transPDSexpanded$ where $\initconfig = (\initpdsstate,\emptyword)$ is the \emph{initial configuration}, $\transPDSrel \subseteq \configPDS \times \PDSrel \times \configPDS$ is the \emph{transition relation}, and $\labelfct$ is the \emph{labeling function} that extends the labeling function of $\PDS$ to configurations as follows: $\labelfct(\pdsstate,\stword) = \labelfct(\pdsstate)$. The transition relation $\transPDSrel$ is defined to capture the informal semantics of internal, call and return transitions --- for any $ \stword \in \stalph^*$, ($\internal$) $(\pdsstate, \stword) \PDSlabel[(\internal,\pdsstate,\emptyword,\pdsstate')] (\pdsstate', \stword)$ iff $(\pdsstate,\pdsstate') \in \PDSrel_{\internal}$; ($\call$) $(\pdsstate, \stword) \PDSlabel[(\call,\pdsstate,\stele,\pdsstate')] (\pdsstate', \stele\stword)$ iff $(\pdsstate, (\pdsstate', \stele)) \in \PDSrel_{\call}$; and ($\return$) $(\pdsstate, \stele\stword) \PDSlabel[(\return,\pdsstate,\stele,\pdsstate')] (\pdsstate', \stword)$ iff $((\pdsstate, \stele), \pdsstate') \in \PDSrel_{\return}$.
	
	A \emph{path} of $\transPDS$ is an infinite sequence of configurations $\exec = \config_0,\config_1,\ldots$ such that for each $i$, $c_i \PDSlabel c_{i+1}$ for some $\s{o} \in \set{\internal,\call,\return}$, $\pdsstate,\pdsstate' \in \pdsstates$ and $\stele \in \stalph \cup \set{\emptyword}$. The path $\exec$ is said to \emph{start} in configuration $\config_0$ (the first configuration in the sequence). We will use $\paths(\transPDS, \config)$ to denote the set of paths of $\transPDS$ starting in the configuration $\config$ and $\paths(\transPDS)$ to denote all paths of $\transPDS$.
	
	We conclude this section by introducing some notation on configurations. For $\config = (\pdsstate,\stword)$, its \emph{stack height} is $|\stword|$, its \emph{control state} is $\state(\config) = \pdsstate$, and its \emph{top of stack symbol} is $\topstk(\config) = \stele \in \stalph$ if $\stword = \stele\stword'$ and is undefined if $\stword = \emptyword$.
	
	\section{\modltl and \rmodltl}
	Checking {\hypltl} and {\hypctl} properties for pushdown systems is undecidable~\cite{model-pommellet}. In this paper, we present a new variant of {\hypctl}  called {\ourlogicFull} ({\modltl} for short), and we will show that checking if a pushdown system satisfies a {\modltl} property is decidable. The complexity of deciding this question for pushdown systems is almost the same as the one for checking if a {\hypctl} property holds on a finite state transition system. {\modltl} is very similar to {\hypctl} in terms of syntax and semantic intent. Recall that a {\hypctl} property typically expresses constraints that must be satisfied by sets of executions of a system. The key innovation in the semantics of {\modltl} (to yield decidable pushdown model checking) is the requirement that executions in a collection satisfying a property be synchronized in terms of their stack operations, i.e., at any step, either all the executions push or all pop or all leave the stack unchanged. This informal requirement will be clearer when the semantics is spelt out. We begin with the syntax of formulas. As in the case of {\hypltl}, Stack-aware Hyper Linear Temporal Logic ({\rmodltl} for short) is defined as a sub-logic of {\modltl}.
	
	\paragraph{Syntax.}
	
			\begin{figure}
			\centering
				\begin{framed}
			
				$\atomprop \in \atomprops,\pathvar\in\pathvars$
				
			\begin{subfigure}[b]{0.40\textwidth}
				\[
				\begin{array}{ccccccccc}
					\spformula & \coleq & \atomprop_\pathvar &\vert& \lnot\spformula &\vert& \spformula \orsymb \spformula &\vert& \next \spformula \\
					 & & &  \vert& \spformula \until \spformula &\vert& \existspi \spformula \\
				\end{array}
				\]
				\subcaption{{\hypctl}}
				\label{sub:hctl2}
			\end{subfigure}
		\hfill
			\begin{subfigure}[b]{0.54\textwidth}
				\[
				\begin{array}{ccccccccccccc}
				    \\
					\ctlformula & \coleq & \apsi &\vert& \epsi &\vert& \lnot \ctlformula &\vert&\ctlformula \orsymb \ctlformula \\ 
					
					\spformula & \coleq & \atomprop_\pathvar &\vert& \lnot\spformula &\vert& \spformula \orsymb \spformula &\vert& 
					\next \spformula &\vert& \spformula \until \spformula &\vert& \existspi \spformula
				\end{array}
				\]
				\subcaption{{\modltl}}
				\label{sub:shctl2}
			\end{subfigure}
			\end{framed}
		\caption{BNF for {\hypctl} and {\modltl}.  {\hypltl} is the set of {\hypctl} formulas $Q_1 \pi_1. \cdots Q_r \pi_r. \psi$ where $Q_i\in \set{\exists,\forall}$, $\psi$ is quantifier free and $\forall\psi=\lnot \exists \lnot \psi$. {\rmodltl} is the set of {\modltl} formulas $\genquant\ Q_1 \pi_1. \cdots Q_r \pi_r. \psi$ where $\genquant\in \set{A,E}$ and $Q_i\in \set{\exists,\forall}$, $\psi$ is quantifier free and $\forall\psi=\lnot \exists \lnot \psi$.}
		\label{fig:hctlvsshctl2}
	\end{figure}

	Let us fix a (finite) set of atomic propositions $\atomprops$, and a set of path variables, $\pathvars$. The BNF grammar for {\modltl} formulas is given in Figure~\ref{fig:hctlvsshctl2}(b). In the BNF grammar,
	$\atomprop \in \atomprops$ is an \emph{atomic proposition}, $\pathvar$ is a \emph{path variable}, $\spformula$ is a \emph{\stprofform}, and $\ctlformula$ is a {\modltl} formula. Like in {\hypctl}, formulas in {\modltl} are built from atomic propositions, path variables, Boolean and modal connectives, and path quantifiers. The syntax is identical to that of {\hypctl} formulas except for the outermost $A$ and $E$ quantifiers that quantify over the different stack access patterns in recursive computations (see \secref{semantics}). Further, following~\cite{templogichyperprop-clarkson,modelchecking-finkbeiner}, we assume that the until operator $\until$ occurs only within the scope of a path quantifier. 
			

	We call formulas of the form $\genquant\spformula$ (where $\genquant \in \set{A,E}$ and $\spformula$ is a {\stprofform}) \emph{{\basicfml} formulas}. Observe that any {\modltl} formula is a Boolean combination of {\basicfml} formulas. A {\modltl} formula $\ctlformula$ is a \emph{sentence} if in every {\basicfml} sub-formula $\genquant\spformula$, $\spformula$ is a sentence, i.e., every path variable appearing in $\spformula$ is quantified. We will assume, without loss of generality, that in any {\stprofform} $\spformula$, all bound path variables in $\spformula$ are renamed to ensure that any path variable is quantified at most once. We will only consider {\modltl} sentences in this paper. The logic {\rmodltl} is a sub-logic of {\modltl}
	and consists of all formulas of the form $\genquant\ Q_1 \pi_1. \cdots Q_r \pi_r. \psi$
	where where $\genquant\in \set{A,E}$ and $Q_i\in \set{\exists,\forall}$, $\psi$ is quantifier free and $\forall\psi=\lnot \exists \lnot \psi$where $\genquant\in \set{A,E}$ and $Q_i\in \set{\exists,\forall}$, $\psi$ is quantifier free and $\forall\psi=\lnot \exists \lnot \psi.$ 
	
	\begin{remark}
	    We have chosen to not have conjunction as an explicit logical operator, to keep our exposition simple. This choice does makes the automata constructions presented here less efficient for formulas involving conjunction. Adding conjunction explicitly does not pose a technical challenge to our setup and our automata constructions can be extended to explictly handle them.
	    
	    In addition to conjunction, we will sometimes use other quantifiers and logical operators to write formulas. Some standard examples include: $\apsi = \lnot E \lnot \psi$; $\ctlformula_1 \andsymb \ctlformula_2 = \lnot (\lnot \ctlformula_1 \orsymb \lnot \ctlformula_2)$, where $\ctlformula_i$ ($i \in \set{1,2}$) is either a {\modltl} or {\stprofadj} formula; $\forall\pathvar.\: \spformula = \lnot \existspi \lnot \spformula$; $\diamondsymb \spformula = \tru\until\spformula$, where $\tru = \atomprop_\pathvar \orsymb \lnot\atomprop_\pathvar$; $\squaresymb\spformula = \lnot\diamondsymb\lnot\spformula$.
	\end{remark}
	
	\subsection{Semantics of \modltl}
	\seclabel{semantics}
	
	
	The syntax of {\stprofform}s is identical to that {\hypctl} formulas. Their semantics will be described in a similar manner, in a context where free path variables in the formula are interpreted as executions of a system. However, we will require that the interpretations of every path variable share a \emph{common} stack access pattern --- hence the term \emph{\stprofadj}. Thus, before defining the semantics, we will define what we mean by the stack access pattern of a path or \emph{stack access pattern } and a \emph{path environment} that assigns an interpretation to path variables. 
	
	For the rest of this section let us fix a pushdown system $\PDS = \PDSexpanded$. A string $\word \in \set{\call,\internal,\return}^*$ is said to be \emph{well matched} if either $\word = \emptyword$ or $\word = \internal$ or $\word = \call\: u\: \return$ or $\word = uv$, where $u, v \in \set{\call,\internal,\return}^*$ are (recursively) well matched. In a string $\stprof \in \set{\call,\internal,\return}^\omega$, $\ith{\stprof}$ is an \emph{unmatched return}, if $\prefix[i+1]{\stprof} = \word\:\return$, where $\word$ is well matched. We are now ready to present the definition of a stack access pattern.
	
	\begin{definition}[Stack access pattern]
		\sloppy A string $\stprof \in \set{\call,\internal,\return}^\omega$ is a \emph{stack access pattern } if the set $\setpred{i \in \natnums}{\ith{\stprof} \mbox{ is an unmatched return}}$ is finite.
		
		For a path $\exec = \config_0 \config_1 \config_2 \cdots \in \paths(\transPDS)$ and stack access pattern $\stprof = o_0 o_1 \cdots$, we say $\exec$ \emph{has stack access pattern } $\stprof$ (denoted $\pathprof(\exec) = \stprof$) if for every $i$:
		\begin{enumerate*}[nosep,label=(\alph*)]
			\item $o_i = \call$ if and only if $\stack(c_{i+1})$ = $\topstk(c_{i+1})\, \stack(c_i)$, 
			\item $o_i = \internal$ if and only if $\stack(c_{i+1}) = \stack (c_i)$, and
			\item $o_i = \return$ if and only if $\stack(c_i) =\topstk(c_i)\, \stack (c_{i+1})$.
		\end{enumerate*}
	\end{definition}
	
	We now present the definition of \emph{path environment} that interprets the free path variables in a {\stprofform} as paths of $\transPDS$ such that they share a common stack access pattern. This plays a key role in defining the semantics of {\modltl}. For a set of path variables $\pathvars$, let $\pathvardagger$ be defined as the set $\pathvars \disjunion \set{\spsymbol}$.
	
	\begin{definition}[Path Environment]
		A \emph{path environment} for pushdown system $\PDS$ over variables $\pathvars$ is function $\pathenv: \pathenvexp$ such that $\pathenv(\spsymbol)$ is a stack access pattern , and for every $\pathvarsele \in \pathvars$, $\pathenv(\pathvarsele) \in \paths(\transPDS)$ with $\pathprof(\pathenv(\pathvarsele)) = \pathenv(\spsymbol)$. When the pushdown system is clear from the context, we will simply refer to it as a path environment over $\pathvars$.
		
		When $\pathvars = \emptyset$, we additionally require that there is a $\exec \in \paths(\transPDS,\initconfig)$ (where $\initconfig$ is the initial configuration of $\transPDS$) such that $\pathprof(\exec) = \pathenv(\spsymbol)$.
	\end{definition}
	
	
	We introduce some notation related to path environments that we will need. Let us fix a path environment $\pathenv$ over variables $\pathvars$. Given a path $\exec \in \paths(\transPDS)$, $\pathenv[\pathvar \mapsto \exec]$ denotes the path environment over $\pathvars \union \set{\pathvar}$ such that $\pathenv[\pathvar \mapsto \exec](\pathvar) = \exec$, and $\pathenv[\pathvar \mapsto \exec](\pathvar') = \pathenv(\pathvar')$, for any $\pathvar' \in \pathvardagger$ with $\pathvar' \neq \pathvar$. Finally, for $i \in \natnums$, $\suffix{\pathenv}$ denotes the \emph{suffix} path environment, where every variable is mapped to the suffix of the path starting at position $i$. More formally, for every $\pathvar' \in \pathvardagger$, $\suffix{\pathenv}(\pathvar') = \suffix{\pathenv(\pathvar')}$.
	
	We now define when a pushdown system $\PDS$ satisfies a {\modltl} sentence $\ctlformula$, denoted $\PDS \satisfy \ctlformula$. The definition of satisfaction of $\ctlformula$ relies on a definition of satisfaction for {\stprofform}s. To inductively to define the semantics of {\stprofform}s, we will interpret free path variables using a path environment. Finally, as in {\hypctl}, it is important to track the most recently quantified path variable because that influences the semantics of $\exists \pi (\cdot)$. Thus satisfaction of {\stprofform}s takes the form $\PDS,\pathenv,\pathvarsele' \satisfy \spformula$, where $\pathvarsele'$ is the most recently quantified path variable, and $\pathenv$ is a path environment over the free variables of $\spformula$. Finally, by convention, we will take $\paths(\transPDS,\pathenv(\spsymbol)(0))$ to mean $\paths(\transPDS,\initconfig)$, where $\initconfig$ is the initial configuration of $\transPDS$~\footnote{We need such a convention because $\pathenv(\spsymbol)(0)$ is not a configuration but an element of the set $\set{\call,\internal,\return}$.}. Below, $\ctlformula, \ctlformula_1,\ctlformula_2$ are {\modltl} sentences, while $\spformula,\spformula_1,\spformula_2$ are {\stprofform}s.
	\[
		\begin{array}{l}
		    \PDS \satisfy \lnot \ctlformula \text{ iff } \PDS \notsatisfy \ctlformula\\
		    \PDS \satisfy \ctlformula_1 \orsymb \ctlformula_2 \text{ iff } \PDS \satisfy \ctlformula_1 \text{ or } \PDS \satisfy \ctlformula_2 \\

			\PDS \satisfy \epsi \text{ iff } \text{for some path environment } \pathenv \text{ over } \emptyset, \PDS, \pathenv, \spsymbol \satisfy \spformula \\
			
			\PDS, \pathenv, \pathvarsele'  \satisfy \atomprop_\pathvar \text{ iff } \atomprop \in \labelfct(\ith[0]{\pathenv(\pathvar)}) \\
			
			\PDS, \pathenv, \pathvarsele' \satisfy \lnot\spformula \text{ iff } \PDS, \pathenv, \pathvarsele' \notsatisfy \spformula \\
			
			\PDS, \pathenv, \pathvarsele' \satisfy \spformula_1 \orsymb \spformula_2 \text{ iff } \PDS, \pathenv, \pathvarsele' \satisfy \spformula_1 \text{ or } \PDS, \pathenv, \pathvarsele' \satisfy \spformula_2 \\
			
			
			\PDS, \pathenv, \pathvarsele' \satisfy \next \spformula \text{ iff } \PDS, \suffix[1]{\pathenv},\pathvarsele' \satisfy \spformula \\
			
			\PDS, \pathenv, \pathvarsele' \satisfy \spformula_1 \until \spformula_2 \text{ iff } \exists i \geq 0 : \PDS,\suffix{\pathenv}, \pathvarsele' \satisfy \spformula_2 \text{ and } \forall j, 0 \leq j < i, \\ 
			\hspace*{3.4cm}\PDS,\suffix[j]{\pathenv}, \pathvarsele' \satisfy \spformula_1 \\
			
			\PDS, \pathenv, \pathvarsele' \satisfy \existspi \spformula \text{ iff } \exists \exec \in \paths(\transPDS, \ith[0]{\pathenv(\pathvarsele')}) \text{with } \pathprof(\exec)=\pathenv(\spsymbol), \text{ such that}\\
			\hspace*{3cm}\text{such that } \PDS,\pathenv[\pathvar \mapsto \exec], \pathvar \satisfy \spformula
		\end{array}
	\]
	
	\fi
	\section{A Decision Procedure for {\modltl}}
	
	Given a pushdown system $\PDS$ and a {\modltl} sentence $\ctlformula$, we present an algorithm that determines if $\PDS \satisfy \ctlformula$. Our approach is similar to the one in~\cite{modelchecking-finkbeiner}. Given a finite state transition system ${\cal K}$ and a {\hypctl} formula $\varphi$, Finkbeiner et. al.~\cite{modelchecking-finkbeiner}, construct an alternating (finite state) B\"{u}chi automaton ${\cal A}_{{\cal K},\varphi},$ {by induction on $\varphi,$} such that an input word $\exec$ is accepted by ${\cal A}_{{\cal K},\varphi}$ if and only if $\exec$ is the encoding of a path environment $\pathenv$ such that ${\cal K}, \pathenv \satisfy \varphi$. Determining if ${\cal K} \satisfy \varphi$ then reduces to checking if ${\cal A}_{{\cal K},\varphi}$ accepts any string.
	
	Extending these ideas to {\modltl} and pushdown systems, requires one to answer two questions: (a) What is an encoding of path environments for {\stprofadj} formulas where path variables are mapped to sequences of configurations (control state + stack)?; (b) Which automata models can capture the collection of path environments satisfying a {\stprofadj} formula with respect to a pushdown system? We encode path environments for {\stprofadj} formulas using strings over a \emph{pushdown alphabet} --- pushdown tags on symbols adds structure that helps encode sequences of configurations. And for automata, we consider automata that process such strings and accept \emph{visibly pushdown languages}. A natural generalization of the approach outlined in~\cite{modelchecking-finkbeiner} would suggest the use of alternating visibly pushdown automata (AVPA) on infinite strings~\cite{vpa-bozzelli}. However, using AVPAs results in an inefficient algorithm. To get a more efficient algorithm, we instead rely on a careful use of \emph{nondeterministic visibly pushdown automata (NVPA)}~\cite{vpl-alur} and \emph{1-way alternating jump automata ($1$-AJA)}~\cite{vpa-bozzelli}.
	The advantage of using NVPA and $1$-AJA can be seen in the case of 
	existential quantification ($\existspi$) which requires converting an alternating automaton to a nondeterministic one~\cite{modelchecking-finkbeiner}: Converting from  $1$-AJA to NVPA  leads to exponential blowup while converting AVPA to NVPA leads to a doubly exponential blowup~\cite{vpa-bozzelli}.
	
	The rest of this section is organized as follows. We begin by introducing the automata models on pushdown alphabets (\secref{vpa-aja}). Next we present our encoding of path environments, and finally our automata constructions that establish the decidability result (\secref{algo}).
	
	\subsection{Automata on Pushdown Alphabets}
	\seclabel{vpa-aja}
	
	\sloppy A \emph{pushdown alphabet} is a finite set $\alphabet$ that is partitioned into three sets $\vpalphexpanded$, where $\alphabet_\call$ is the set of \emph{call symbols}, $\alphabet_\internal$ is the set of \emph{internal symbols}, and $\alphabet_\return$ is the set of \emph{return symbols}. Automata models processing strings over a pushdown alphabet are restricted to perform certain types of transitions based on whether the read symbol is a call, internal, or return symbol. We introduce, informally, two such automata models next. Precise definition and its semantics can be found in \ifdefined\submittedshort \appref{nvpa} and \appref{aja}.
	\else 
	the detailed version of this paper~\cite{fullpaper}. 
	\fi
	
	\paragraph{Nondeterministic Visibly Pushdown B\"{u}chi Automata.}
	A \emph{nondeterministic visibly pushdown automaton (NVPA)}~\cite{vpl-alur} is like a pushdown system. It has finitely many control states and uses an unbounded stack for storage. However, unlike a pushdown system, it is an automaton that processes an infinite sequence of input symbols from a pushdown alphabet $\alphabet = \vpalphexpanded$. Transitions are constrained to conform to pushdown alphabet --- whenever a $\alphabet_\call$ symbol is read, a symbol onto the stack, whenever a $\alphabet_\return$ symbol is read, the top stack symbol is popped, and whenever $\alphabet_\internal$ symbol is read, the stack is unchanged. 
	
	\paragraph{1-way Alternating Jump Automata.}
	Our second automaton model is \emph{1-way Alternating Parity Jump Automata (1-AJA)}~\cite{vpa-bozzelli}. 1-AJA are computationally equivalent to NVPAs (i.e., accept the same class of languages) but provide greater flexibility in describing algorithms. 1-AJAs are alternating automata, which means that they can define acceptance based on multiple runs of the machine on an input word. Though they are finite state machines with no auxiliary storage, their ability to spawn a computation thread that jumps to a future portion of the input string on reading a symbol, allows them to have the same computational power as a more conventional machine with storage (like NVPAs). 
	
	We present some useful properties of NVPA and 1-AJA. The two models are equi-expressive with the size of automata constructed by the translation known.
	\begin{theorem}[\cite{vpa-bozzelli}]
		\thmlabel{aja-vpa}
		For any NVPA $\autoVPA$ of size $n$, there is a 1-AJA $\auto_\autoVPA$ of size $O(n^2)$, such that $\lang(\auto_\autoVPA) = \lang(\autoVPA)$. Conversely, for any 1-AJA $\auto$ of size $n$, there is a NVPA $\autoVPA_\auto$ of size $2^{O(n)}$, such that $\lang(\autoVPA_\auto) = \lang(\auto)$. 
		Constructions can be carried out in time proportional to the size of the resulting automaton.
	\end{theorem}
	
	Both 1-AJA and NVPAs are closed for language operations like complementation, union and prefixing. 
	We also recall the following result.
	%
	\begin{theorem}
		\thmlabel{nvpa-emptiness}
		(\cite{vpl-alur}) For NVPAs, the emptiness problem is {\ptime}-complete.
	\end{theorem}
	
	\subsection{Algorithm for {\modltl}}
	\seclabel{algo}
	
	Let us fix a pushdown system $\PDS = \PDSexpanded$ and a {\modltl} sentence $\ctlformula$. Our goal is to decide if $\PDS \satisfy \ctlformula$. We will reduce this problem to checking the emptiness of multiple NVPAs (\thmref{nvpa-emptiness}). Our approach is similar to~\cite{modelchecking-finkbeiner} --- for each {\stprofadj} sub-formula $\spformula$ (not necessarily sentence) of $\ctlformula$, we will compositionally construct an automaton that accepts the path environments satisfying $\spformula$. Path environments will be encoded by strings over pushdown alphabets as follows.
	
	For a path $\exec = \config_0 \config_{1} \config_2\cdots$ of $\transPDS$, the \emph{trace} of $\exec$, denoted $\tr(\exec)$, is the (unique) sequence $(\s{o}_0,\ctrlstate_0,\stele_0,\ctrlstate_1)(\s{o}_1,\ctrlstate_1,\stele_1,\ctrlstate_2)\cdots$ such that for every $i \in \natnums$, $\config_i \PDSlabel[(\s{o}_i,\ctrlstate_i,\stele_i,\ctrlstate_{i+1})] \config_{i+1}$ where $\s{o}_i \in \set{\call,\internal,\return}$, $\ctrlstate_i,\ctrlstate_{i+1} \in \ctrlstates$, and $\stele_i \in \stalph \cup \set{\emptyword}$~\footnote{Observe that even when $\exec$ is not a path in $\transPDS$ (i.e., corresponds to an actual sequence of transitions of $\PDS$), the trace of $\exec$ is uniquely defined 
	as long as stacks of successive configurations of $\exec$ can be obtained by leaving the stack unchanged, or pushing/popping one symbol.}. 
	
	While $\tr(\exec)$ is uniquely determined by the path $\exec$, the converse is not true --- different paths may have the same trace. To see this, consider the following example. For configuration $\config$ and $\gamma \in \stalph^*$, let $\gamma(\config)$ denote the configuration $(\state(\config),\stack(\config)\gamma)$, i.e., the configuration with the same control state, but with stack containing the symbols in $\gamma$ at the bottom. Observe that, for any $\gamma \in \stalph^*$, if $\exec = \config_0\config_1\config_2\cdot$ is a path then so is $\gamma(\exec) = \gamma(\config_0)\gamma(\config_1)\gamma(\config_2)\cdots$. Additionally, $\tr(\exec) = \tr(\gamma(\exec))$. Two paths $\exec_1$ and $\exec_2$ of $\transPDS$ will be said to be {\pathequi} if $\tr(\exec_1) = \tr(\exec_2)$ and will be denoted as $\exec_1 \treq \exec_2$. Observe that {\pathequi} paths have the same stack access pattern , i.e. if $\exec_1 \treq \exec_2$ then $\pathprof(\exec_1) = \pathprof(\exec_2)$. The semantics of {\modltl} doesn't distinguish between {\pathequi} paths.
	\begin{proposition}
		\proplabel{eq-path-envs}
		Let $\varphi$ be a {\stprofform} with $\pathvars$ as the set of free path variables. Let $\pathenv_1$ and $\pathenv_2$ be two path environments such that for every $\pi \in \pathvars$, $\pathenv_1(\pi) \treq \pathenv_2(\pi)$. Then, $\PDS,\pathenv_1,\pathvarsele \satisfy \varphi$ if and only if $\PDS,\pathenv_2,\pathvarsele \satisfy \varphi$. 
	\end{proposition}
	
	The proof of \propref{eq-path-envs} follows by induction on {\stprofform}s. \propref{eq-path-envs} establishes that the set of path environments satisfying a {\stprofform} is a union of equivalence classes with respect to path equivalence. Thus, instead of constructing automata that accept path environments, we will construct automata that accept mappings from path variables to traces of paths. For $m \in \natnums$, let $\alphabet[m] = \alphabet[m]_\call \disjunion \alphabet[m]_\internal \disjunion \alphabet[m]_\return$ be the pushdown alphabet where $\alphabet[m]_\call = \set{\call} \times \pdsstates^m \times \stalph^m$, $\alphabet[m]_\internal = \set{\internal} \times \pdsstates^m \times \set{\emptyword}^m$, and $\alphabet[m]_\return = \set{\return} \times \pdsstates^m \times \stalph^m$. Observe $\alphabet[0]$ is (essentially) the set $\set{\internal,\call,\return}$.
	\begin{definition}[Encoding Path Environments]
		\deflabel{encode}
		Consider a set of $m$ path variables $\pathvars = \set{\pathvarsele_1,\pathvarsele_2,\ldots \pathvarsele_m}$. A string $\word \in \alphabet[m]^\omega$ where for any $j \in \natnums$, $\ith[j]{\word} = (\s{o}_j, (\pdsstate^j_1,\pdsstate^j_2,\ldots \pdsstate^j_m), (\stele^j_1,\stele^j_2,\ldots \stele^j_m))$ encodes all path environments $\pathenv$ such that
		\[
		\begin{array}{l}
			\pathenv(\spsymbol) = \s{o}_0\s{o}_1\s{o}_2\cdots  \s{o}_j\cdots \\
			\tr(\pathenv(\pathvarsele_i)) = (\s{o}_0,\pdsstate^0_i,\stele^0_i,\pdsstate^1_i)(\s{o}_1,\pdsstate^1_i,\stele^1_i,\pdsstate^2_i)\cdots
		\end{array}
		\]
		for any $i \in \set{1,2,\ldots m}$. The string encoding a path environment $\pathenv$ is denoted as $\enc(\pathenv)$ ($=\word$, in this case).
	\end{definition}
	
	Based on the definitions, the following observation about traces and encodings can be concluded.
	\begin{proposition}
		\proplabel{enc-suffix}
		For any path $\exec \in \paths(\transPDS)$ and $i \in \natnums$, $\tr(\suffix{\exec}) = \suffix{\tr(\exec)}$. For any path environment $\pathenv$ and $i \in \natnums$, $\enc(\suffix{\pathenv}) = \suffix{\enc(\pathenv)}$.
	\end{proposition}
	
	The encoding of path environments as strings over $\alphabet[m]$ (for an appropriate value of $m$) is used in our decision procedure, which compositionally constructs automata that accept path environments satisfying each {\stprofform}. The size of our constructed automata, like in~\cite{modelchecking-finkbeiner}, will be tower of exponentials that depends on the {\emph{formula complexity}} of the {\stprofform} $\varphi$. 
	\begin{definition}[{{Formula Complexity}}]
		\deflabel{alt-depth}
		The {\emph{formula complexity}} of a {\modltl} formula $\varphi$, denoted $\alt(\varphi)$, is inductively defined as follows. Let $\odd: \natnums \to \natnums$ be the function that maps a number $n$ to the smallest odd number $\geq n$, i.e., $\odd(n) = n$ if $n$ is odd and $\odd(n) = n+1$ if $n$ is even. Similarly, $\even: \natnums \to \natnums$ maps $n$ to the smallest even number $\geq n$, i.e., $\even(n) = \odd(n+1) - 1$. Below $\spformula_1,\spformula_2$ denote {\stprofform}s, and $\ctlformula_1,\ctlformula_2$ denote {\modltl} sentences.
		\[
		\begin{array}{lcl}
			\multicolumn{3}{c}{\alt(\atomprop_\pathvar) = 0 \quad\qquad \alt(\lnot\spformula_1) = \even(\alt(\spformula_1)) \quad\qquad \alt(\next \spformula_1) = \alt(\spformula_1)}\\
			\alt(\spformula_1 \orsymb \spformula_2) = \max (\alt(\spformula_1), \alt(\spformula_2)) & \quad & \alt(\spformula_1 \until \spformula_2) = \even(\max (\alt(\spformula_1),\alt(\spformula_2)))\\ 
			\alt(\existspi \spformula_1) = \odd(\alt(\spformula_1)) & \quad & \alt(E \spformula_1) = \odd(\alt(\spformula_1))\\
			\alt(\lnot\ctlformula_1) = \alt(\ctlformula_1) & \quad & \alt(\ctlformula_1 \orsymb \ctlformula_2) = \max (\alt(\ctlformula_1),\alt(\ctlformula_2))\\
		\end{array}
		\]
	\end{definition}
	Observe the difference in the definition of $\alt(\lnot\ctlformula_1)$ and $\alt(\lnot\spformula_1)$; for $\lnot\ctlformula_1$ there is no change in {formula complexity},
 while for $\lnot\spformula_1$ we move to the next even level.
	
	Our main technical lemma is a compositional construction of an automaton for {\stprofform}s $\spformula$. Depending on the parity of $\alt(\spformula)$, the automaton we construct will either be a 1-AJA or a NVPA. Before presenting this lemma, we define a function that is a tower of exponentials. For $c,k,n \in \natnums$, the value $g_c(k,n)$ is defined inductively on $k$ as follows: 	$g_c(0,n) = cn\log n$, and $g_c(k+1,n) = 2^{g_c(k,n)}.$
	We use $g_{O(1)}(k,n)$ to denote the family of functions $\setpred{g_c(k,n)}{c \in \natnums}$.
	\begin{lemma}
		\lemlabel{main}
		Consider pushdown system $\PDS = \PDSexpanded$ and {\modltl} sentence $\ctlformula$. Let $\spformula$ be a {\stprofadj} subformula of $\ctlformula$ with free path variables in the set $\pathvars = \set{\pathvarsele_1,\ldots\pathvarsele_m}$ for $m \in \natnums$. We assume, without loss of generality, that the variables $\pathvarsele_1,\ldots\pathvarsele_m$ are in the order in which they are quantified in $\ctlformula$ with $\pathvarsele_m$ being the first free variable of $\spformula$ that will be quantified in the context $\ctlformula$. In addition, we assume that the size of both $\spformula$ and $\PDS$ is bounded by $n$. There is an automaton $\auto_\spformula$ over pushdown alphabet $\alphabet[m]$ such that for any path environment $\pathenv$ over $\pathvars$,
		\[
		\PDS,\pathenv,\pathvarsele_m \satisfy \spformula \text{ if and only if } \enc(\pathenv) \in \lang(\auto_\spformula).~\footnote{When $m = 0$, we take $\pathvarsele_m$ to be $\spsymbol$.}
		\]
		The automaton $\auto_\spformula$ is a NVPA if $\alt(\spformula)$ is odd, and a 1-AJA if $\alt(\spformula)$ is even. The size of $\auto_\spformula$ is at most $g_{O(1)}(\ceil{\frac{\alt(\spformula)}{2}},n)$\footnote{When the size of the specification $\spformula$  is considered constant, the size of $\auto_\spformula$ is at most $g_{O(1)}(\ceil{\frac{\alt(\spformula)}{2}}-1,n)$ }.
	\end{lemma}
	
	Before presenting the proof of \lemref{main}, we would like to highlight a subtlety about its statement. The result guarantees that for \emph{valid} path environments $\pathenv$, encoding $\enc(\pathenv)$ is accepted by $\auto_\spformula$ if and only if $\pathenv$ satisfies $\spformula$. It says nothing about path environments that are not valid. In particular, there may be functions that map path variables to traces that do not correspond to actual paths of $\transPDS$, but which are nonetheless accepted by $\auto_\spformula$. Notice, however, when $\spformula = \existspi\spformula_1$ is a {\stprofadj} sentence, a string over $\set{\call,\internal,\return}$ will, by conditions guaranteed in \lemref{main}, be accepted if and only if it corresponds to a stack access pattern of a path from the initial state that satisfies $\existspi\spformula_1$.
	
	\begin{proof}[Sketch of \lemref{main}]
		Our construction of $\auto_\spformula$ will proceed inductively. The type of automaton constructed will be consistent with the parity of $\alt(\spformula)$, i.e., an NVPA if $\alt(\varphi)$ is odd and a 1-AJA if $\alt(\spformula)$ is even. We sketch the main ideas here, with the full proof 
		\ifdefined\submittedshort
		available in~\appref{proof-main}.
		\else
		in~\cite{fullpaper}.
		\fi
		
		For $\atomprop_\pathvar$, $\lnot\spformula_1$, $\spformula_1 \orsymb \spformula_2$, and $\next\spformula_1$, the construction essentially proceeds by converting $\auto_{\spformula_i}$ ($i \in \set{1,2}$) if needed, into the type (NVPA or 1-AJA) of the target automaton using \thmref{aja-vpa}, and then using standard closure properties to combine them to get the desired automaton. In case of $\spformula = \spformula_1\until\spformula_2$, we first convert (if needed) $\auto_{\spformula_i}$ ($i \in \set{1,2}$) into a 1-AJA. At each step, the automaton for $\spformula$ will choose to either run $\auto_{\spformula_2}$, or run $\auto_{\spformula_1}$ \emph{and} restart itself. Correctness relies on the fact that our encoding for path environments satisfies \propref{enc-suffix}.
	
		The most interesting case is that of $\spformula = \existspi\spformula_1$. We will first convert (if needed) the automaton for $\spformula_1$ into a NVPA $\auto_1$. The automaton for $\spformula$ will essentially guess the encoding of a path that is consistent with the transitions of $\PDS$, and check if assigning the guessed path to variable $\pathvar$ satisfies $\spformula_1$ by running the automaton $\auto_1$. The additional requirement we have is that the guessed path start at the \emph{same configuration} as the current configuration of the path assigned to variable $\pathvarsele_m$ which introduces some subtle challenges. In order to be able to guess a path, $\auto_\spformula$ will keep track of $\PDS$'s control state in its control state, and use its stack to track $\PDS$'s stack operations along the guessed path. Since the stacks of all paths are synchronized, it makes it possible for $\auto_\spformula$ to use its (single stack) to track the stack of both $\PDS$ and the stack of $\auto_1$.\qed
	\end{proof}

	Using \lemref{main}, we can establish the main result of this section. 
	\begin{theorem}
	    \thmlabel{main}
	    Given a $\PDS = \PDSexpanded$ and a {\modltl} sentence $\ctlformula$, the problem of determining if $\PDS \satisfy \ctlformula$ is in $\cup_c \dtime{g_c(\ceil{\frac{\alt(\ctlformula)}{2}},n)}$, where $n$ is a bound on the size of $\PDS$ and $\ctlformula$.
	\end{theorem}
	
	\begin{proof}
	    Recall that a {\modltl} sentence is a Boolean combination of formulas of the form $E\spformula$, where $\spformula$ is a {\stprofadj} sentence. Results on whether $\PDS \satisfy E\spformula$ for each such subformula can be combined to determine whether $\PDS \satisfy \ctlformula$. Given this, the time to determine if $\PDS \satisfy \ctlformula$ is at most the time to decide if $\PDS$ satisfies each subformula of the form $E \spformula$ plus $O(n)$ (to compute the Boolean combination of these results). 
	    Next, recall that the construction in \lemref{main} ensures that for a {\stprofadj} sentence of the form $\existspi \spformula$, $\lang(\auto_{\existspi \spformula})$ consists exactly of strings in $\set{\call,\internal,\return}^\omega$ that encode a path environment over $\emptyset$ that satisfy $\existspi\spformula$.
	    
	    Consider a {\modltl} sentence $E\spformula$. Let $\pathvar$ be a path variable that does not appear in the sentence $\spformula$. Based on the semantics of {\modltl} the following observation holds: $\PDS \satisfy E\spformula$ if and only if for some path environment $\pathenv$ over $\emptyset$, $\PDS, \pathenv, \spsymbol \satisfy \existspi\spformula$. Which is equivalent to saying that $\PDS \satisfy E\spformula$ if and only if $\lang(\auto_{\existspi\spformula}) \neq \emptyset$. Since $\alt(E\spformula) = \alt(\existspi\spformula)$, and the emptiness problem of NVPA can be decided in polynomial time (\thmref{nvpa-emptiness}), our theorem follows.\qed
	\end{proof}
	
	\section{Lower Bound}
	\seclabel{lower-bound}
	
	In this section, we establish a lower bound for the problem of model checking {\modltl} sentences against pushdown systems. Our proof establishes a hardness result for the {\rmodltl} sub-fragment of {\modltl}. 
	Before presenting this lower bound, we introduce the function $h_c(\cdot,\cdot)$, which is another tower of exponentials, inductively defined as follows: $	h_c(0,n) = n$, and $h_c(k+1,n) = h_c(k,n)\cdot c^{h_c(k,n)}.$
	
	\begin{theorem}
	    \thmlabel{lower-bound}
	    Let $\PDS$ be a pushdown system and $\ctlformula$ be a {\rmodltl} sentence such that the sizes of both $\PDS$ and $\ctlformula$ is bounded by $n$ and $\alt(\ctlformula) = 2k-1$ for some $k \in \natnums$. The problem of checking if $\PDS \satisfy \ctlformula$ is $\dtime{h_c(k,n)}$-hard, for every $c \in \natnums$.
	\end{theorem}
	
	\begin{proof}[Sketch]
	    We sketch the main intuitions behind the proof. To highlight the novelties of this proof, it is useful to recall how $\textsf{NSPACE}(h_c(k-1,n))$-hardness for {\hypltl} model checking is proved~\cite{templogichyperprop-clarkson}. The idea is to reduce the language of a nondeterministic $h_c(k-1,n)$ space bounded machine $M$ to the model checking problem by constructing a finite state transition system that guesses a run of $M$, and a {\hypltl} formula that checks if the path is a valid accepting run.
	    
	    To get the stricter bound of $\dtime{h_c(k,n)}$, we use the fact that we are checking pushdown systems. The stack of the pushdown system can be used to guess a \emph{tree}, as opposed to a simple trace. Therefore, we reduce a $h_c(k-1,n)$ space bounded \emph{alternating} Turing machine, instead of a nondeterministic machine. Since $\aspace{f(n)} = \dtime{2^{O(f(n))}}$ for $f(n) \geq \log n$, the theorem will follow if the reduction succeeds.
	
	    Recall that a run of an alternating Turing machine $M$ is a rooted, labeled tree, where vertices are labeled by configurations of $M$ in a manner that is consistent with the transition function of $M$. To faithfully encode a tree as a sequence of symbols, we record the DFS traversal of the tree, making explicit the stack operations performed during such a traversal. Consider a labeled, rooted tree $T$ with root $r$ whose label is $\ell(r)$ with $T_1$ as a the left sub-tree and $T_2$ as the right sub-tree. The DFS traversal of $T$ will push $\ell(r)$, traverse $T_1$ recursively, pop $\ell(r)$, push $\ell(r)$, traverse $T_2$, and then pop $\ell(r)$. We will use such a DFS traversal to guess and encode runs of $M$. Popping and pushing $\ell(r)$ between the traversals of $T_1$ and $T_2$ may seem redundant. Why not simply do nothing between the traversals of $T_1$ and $T_2$? For $T$ to be a valid run of $M$, the configuration labeling of the root of $T_2$ must be the result of taking one step from $\ell(r)$. Such checks will be encoded in our {\rmodltl} sentence, and for that to be possible, we need successive configurations of $M$ to be consecutive in the string encoding.
	    
	    To highlight some additional consistency checks, let us continue with our example tree $T$ from the previous paragraph. For a string to be a correct encoding of $T$, it is necessary that the string pushed before the traversal of $T_i$ ($i \in \set{1,2}$) be the same as the string popped after the traversal. This can be ensured by the pushdown system by actually pushing and popping those symbols. In addition, the string popped after $T_1$'s traversal must be the same as the string pushed before $T_2$'s traversal. Neither the stack nor the finite control of the pushdown system can be used to ensure this. Instead this must be checked by the {\rmodltl} sentence we construct. But the symbols while popping $\ell(r)$ will be in reverse order of the symbols being pushed, and it is challenging to perform this check in the formula. To overcome this, we push/pop the label \emph{and its reverse} at the same time. This ensures that if we want to check if a string pushed is the same as a string that was just popped, then we can check for string \emph{equality}, and this check is easier to do using formulas in {\rmodltl}. Additional checks to ensure that the tree encodes a valid accepting run are performed by the {\rmodltl} sentence using ideas from~\cite{qptl-sistla}. Full details can be found 
	    \ifdefined\submittedshort
	      in~\appref{lower-bound-proof}. 
	    \else in~\cite{fullpaper}.
	    \fi 
	    \qed
	\end{proof}

\section{Conclusions}

In this paper, we introduced a branching time temporal logic {\modltl} that can be used to specify synchronous hyperproperties for recursive programs modeled as pushdown systems. The primary difference from the standard branching time logic {\hypctl} for synchronous hyperproperties is that {\modltl} considers a restricted class of hyperproperties, namely, those that relate only executions that the same stack access pattern. We call such hyperproperties stack-aware hyperproperties. We showed that the problem of model checking pushdown systems {\modltl} specifications is decidable, and characterized its complexity. We also showed how this result can potentially be used to aid security verification.

		\bibliographystyle{splncs04}
	\bibliography{references.bib}

\ifdefined\submittedshort
	\appendix

		
	
\section{Observational determinism when the call stack size is visible to the attacker}
 \label{app:observationaldeterminism} 

 	Observational determinism~\cite{obs-mclean,obs-myers} states that any two executions that have the same low-level initial inputs must have the same low-level output observations. Observational determinism is a hypersafety property~\cite{templogichyperprop-clarkson,hyperprop-clarkson}.
 	As~\cite{templogichyperprop-clarkson} shows, observational determinism is also expressible in {\hypltl} using the formula:
	$\mathsf{OD} \stackrel {\mbox{def}}{=}\forallpi \forall\pathvarprime.  (\pi[0] \equiv_{L,in} \pi'[0]) \impsymb  \pi \equiv_{L,out} \pi'.$ Here $\equiv_{L,in}$ and $\equiv_{L,out}$ express the fact that $\pi$ and $\pi'$
	have the same low-security inputs and outputs respectively.

 In order to see how we can verify observational determinism when call stack sizes are observable, let us consider the simple case when the call stack size is the \emph{only} low-security observation. Let $\PDS$ be the pushdown automaton corresponding to a program. We can assume by loss of generality that the states of $\PDS$ encode if the stack is empty, and that pop transitions occur only in states where the stack is empty.\footnote{{Essentially}, initially the stack is taken to be empty. When a symbol is pushed onto an empty stack, it is \lq\lq annotated {\rq\rq} with information that it is the bottom of the stack, and the state remembers that stack is non-empty. When an \lq\lq annotated {\rq\rq} symbol is popped, the stack now remembers that the stack is empty. }  Let $\Delta \PDS$ be the pushdown automaton obtained from $\PDS$ which behaves like $\PDS$, except that it has additional nondeterministic transitions as follows. $\Delta P$ has three new  states $q_\call,$ $q_\internal$ and $q_\return.$ We also assume there are four new propositions: $\mathsf{new}$,  $\call$, $\internal$ and $\return. $   The proposition $\mathsf{new}$ is true in $q_\call, q_\internal$ and $q_\return$ only. The only transitions in the new states are self-loops that do not change the stack size.  	 Whenever $\PDS$ has a push/internal/pop  transition from a state $q$ to $q'$ in $\PDS$, we add an internal nondeterministic transition from $q$ to $q_\call$/$q_\internal$/$q_\return$  that does not change the stack size. Furthermore, the proposition $\call$/$\internal$/$\return$
 	 is true in $q$  in that case. Consider the formula
 	$$\begin{array}{l}
	   A\, \forallpi \forallpi' (\pi[0] \equiv_{L,in} \pi'[0]) \impsymb \\
 	    \hspace*{2cm}((\mathsf{new}_\pi \orsymb \mathsf{new}_{\pi'}) \release \\
 	     \hspace*{3cm } ((\call_{\pi} \Rightarrow \call_{\pi'}\andsymb \lnot \internal_{\pi'} \andsymb \lnot \return_{\pi'}) \andsymb \\
 	     \hspace*{3.8cm } (\internal_\pi  \Rightarrow \internal_{\pi'} \andsymb \lnot \call_{\pi'} \andsymb \lnot \return_{\pi'}) \andsymb \\
 	       \hspace*{4.6cm } (\return_\pi \Rightarrow \return_{\pi'} \andsymb \lnot \call_{\pi'} \andsymb \lnot \internal_{\pi'})))
	   \end{array}$$
	where $\release$ is the dual of $\until$. More precisely, $\psi \release \psi'$ is $\lnot (\lnot \psi') \until (\lnot \psi).$
	
 	Observe that the above formula is not satisfied by the pushdown system $\Delta \PDS$ if and only if there are two \emph{finite} executions $\exec_1$ and $\exec_2$   of $\PDS$ leading to states $q_1$ and $q_2$ such that 
	\begin{enumerate*}[nosep,label=(\alph*)]
	\item the low-level inputs in $\exec_1$ and $\exec_2$ are the same, 
	\item $\exec_1$ and $\exec_2$ have the same stack access pattern, and \item  a push/internal/pop can be executed from only one of the states $q_1$ and $q_2$, but not the other. 
 	\end{enumerate*}
	
	\section{Nondeterministic Visibly Pushdown Automata (NVPA)}
	\applabel{nvpa}

	A \emph{nondeterministic visibly pushdown automaton (NVPA)}~\cite{vpl-alur} is like a pushdown system in that it has finitely many control states and uses an unbounded stack for storage. However, unlike a pushdown system, it is an automaton that processes an infinite sequence of input symbols from a pushdown alphabet $\alphabet = \vpalphexpanded$. Furthermore, transitions are constrained to conform to pushdown alphabet --- whenever the automaton reads from $\alphabet_\call$, it pushes a symbol onto its stack, whenever it reads from $\alphabet_\return$, it pops its top stack symbol, and whenever it reads from $\alphabet_\internal$, it leaves its stack unchanged.
	\begin{definition}[NVPA]
		A \emph{nondeterministic visibly pushdown B\"{u}chi automaton (NVPA)} is a tuple $\autoVPA = \autoexpandedVPA$ where, $\ctrlstates$ is a finite set of control states, $\initctrlstate \in \ctrlstates$ is the initial control state, $\alphabet = \vpalphexpanded$ is a pushdown alphabet that is used to encode inputs, $\stalph$ is a finite set called the stack alphabet, $\stbot \not\in \stalph$ is the bottom of stack symbol, $\acceptstates \subseteq \ctrlstates$ is the set of accepting states, and $\autotransVPA$ is the transition relation. $\autotransVPA$ will be assumed to be partitioned into three as $\autotransexpandedVPA$ where $\autotransVPA_\call \subseteq (\ctrlstates \times \alphabet_\call \times \ctrlstates \times \stalph)$ is the transition on call symbols, $\autotransVPA_{\internal} \subseteq (\ctrlstates \times \alphabet_\internal \times \ctrlstates)$ is the transition on internal symbols, and $\autotransVPA_{\return} \subseteq (\ctrlstates \times \alphabet_\return \times (\stalph \union {\stbot}) \times \ctrlstates)$ is the transition on return symbols.
	\end{definition}
	Let us fix an NVPA $\autoVPA = \autoexpandedVPA$. A \emph{configuration} of $\autoVPA$, like in the case of a pushdown system, is a pair of the form $(\ctrlstate,\stword\stbot)$ where $\ctrlstate \in \ctrlstates$ and $\stword \in \stalph^*$. The functions $\state(\cdot)$, $\stack(\cdot)$, and $\topstk(\cdot)$ are defined in the same manner as for pushdown system configurations. A \emph{run} of $\autoVPA$ on an input string $\word \in \alphabet^\omega$ is an infinite sequence of configurations $\config_0,\config_1,\ldots$ such that $\config_0 = (\initctrlstate,\stbot)$ is the \emph{initial configuration} of $\autoVPA$, and successive configurations are consistent with the input symbol read and the transition relation of $\autoVPA$. More precisely, for every $i \in \natnums$, (a) if $\ith{\word} \in \alphabet_\call$ then $(\state(\config_i),\ith{\word},\state(\config_{i+1}),\topstk(\config_{i+1})) \in \autotransVPA_\call$ and $\stack(\config_{i+1}) = \topstk(\config_{i+1})\stack(\config_i)$; (b) if $\ith{\word} \in \alphabet_\internal$ then $(\state(\config_i),\ith{\word},\state(\config_{i+1})) \in \autotransVPA_\internal$ and $\stack(\config_{i+1}) = \stack(\config_i)$; and (c) if $\ith{\word} \in \alphabet_\return$ then $(\state(\config_i),\ith{\word},\topstk(\config_i),\state(\config_{i+1})) \in \autotransVPA_\return$ and additionally, $\stack(\config_{i+1}) = \stack(\config_i) = \stbot$ if $\topstk(\config_i) = \stbot$ and $\stack(\config_i) = \topstk(\config_i)\stack(\config_{i+1})$ if $\topstk(\config_i) \neq \stbot$. It is worth observing that transitions on call symbols push a symbol onto the stack, transitions on internal symbols leave the stack unchanged, and transitions on return symbols typically pop the top of the stack unless the stack is empty (i.e. $= \stbot$) in which case they leave the stack unchanged.
	
	A run $\config_0,\config_1,\cdots$ on input $\word$ is said to be \emph{accepting} if it satisfies the B\"{u}chi acceptance condition, i.e., for some $\ctrlstate \in \acceptstates$ there are infinitely many $i$ such that $\state(\config_i) = \ctrlstate$. Finally, the \emph{language accepted} by NVPA $\autoVPA$, denoted $\lang(\autoVPA)$, is
	\[
	\lang(\autoVPA) = \setpred{\word \in \alphabet^\omega}{\autoVPA \text{ has an accepting run on } \word}.
	\]
	
	
	\section{1-way Alternating Jump Automata (1-AJA)}
	\applabel{aja}
	
	Our second model of an automaton with strings over a pushdown alphabet, is \emph{1-way Alternating Parity Jump Automata (1-AJA)}~\cite{vpa-bozzelli}. 1-AJA are computationally equivalent to NVPAs (i.e., accept the same class of languages) but provide greater flexibility in describing algorithms. 1-AJAs are alternating automata, which means that they can define acceptance based on multiple runs of the machine on an input word. Though they are finite state machines with no auxiliary storage, their ability to spawn a computation thread that jumps to a future portion of the input string on reading a symbol, allows them to have the same computational power as a more conventional machine with storage (like NVPAs). We present this model after some necessary definitions.
	
	Transitions of an alternating automata identify subsets of next steps that must all be accepting for the machine to accept an input. We describe these subsets of next steps is using Boolean functions. For a set $X$, let $\boolsymb^+(X)$ denote the set of \emph{positive Boolean expressions} built using elements of $X$ as propositions, i.e., they consist of propositional logic formulas built using $\tru$, $\fls$, $X$, $\andsymb$ and $\orsymb$. Given $\varphi \in \boolsymb^+(X)$ and $A \subseteq X$, we say $A \satisfy \varphi$ if $\varphi$ evaluates to true under the valuation that assigns true to the elements of $A$ and false to everything else. The \emph{dual} of a formula $\varphi \in \boolsymb^+(X)$, denoted $\dual{\varphi}$, is the formula obtained from $\varphi$ by replacing $\tru$ by $\fls$, $\fls$ by $\tru$, $\orsymb$ by $\andsymb$, and $\andsymb$ by $\orsymb$. More precisely, $\dual{\cdot}$ can be defined inductively as follows, where $x \in X$.
	\[
	\begin{array}{ccccc}
		\dual{\tru} = \fls & \quad & \dual{\fls} = \tru & \quad & \dual{x} = x\\
		\dual{\varphi \andsymb \psi} = \varphi \orsymb \psi & \quad & \dual{\varphi \orsymb \psi} = \varphi \andsymb \psi
	\end{array}
	\]
	
	A 1-AJA is a finite state automaton that reads an input over a pushdown alphabet. In other words, on reading an input symbol, a thread of computation changes its control state based on its current state and symbol read. However, one of the novel features of a 1-AJA is its ability to \emph{jump} when it reads a call symbol. Thus, a transition of a 1-AJA not only specifies what the next state is, but also what the next symbol to be read is. When the symbol read is either an internal symbol or a return symbol, the symbol to be read next is always the one immediately after; this is denoted as $\nxt$. However, when the symbol read is a call symbol, the automaton can choose to either read the next symbol or to read the \emph{matching return symbol} next; $\anxt$ is used to indicate that the automaton should read the matching return symbol next. We now have all the notation necessary to define a 1-AJA formally.
	\begin{definition}[1-AJA]
		A \emph{1-way Alternating Parity Jump Automaton (1-AJA)} is a tuple $\auto = \autoexpanded$ where, $\ctrlstates$ is a finite set of control states, $\initctrlstate \in \ctrlstates$ is the initial control state, $\alphabet = \vpalphexpanded$ is a pushdown alphabet that is used to encode inputs, $\autotrans : (\ctrlstates \times \alphabet) \to \boolsymb^+(\ctrlstates \times \set{\nxt,\anxt})$ is the transition relation with the restriction that for any $\ctrlstate \in \ctrlstates$ and $\alphele \in \alphabet_\internal \cup \alphabet_\return$, $\autotrans(\ctrlstate,\alphele) \in \boolsymb^+(\ctrlstates \times \set{\nxt})$, and $\parityfct: \ctrlstates \to \natnums$ is the parity function.
	\end{definition}
	
	In order to define a run of 1-AJA on an input string, we need to introduce some notation. Let us fix a pushdown alphabet $\alphabet = \vpalphexpanded$. We can extend the notion of well matched strings to $\alphabet^*$ as follows. A string $z \in \alphabet^*$ is said to be \emph{well matched} if either $z = \emptyword$, or $z \in \alphabet_\internal$, or $z = c\: u\: r$, or $z = uv$, where $c \in \alphabet_\call$, $r \in \alphabet_\return$, and $u, v \in \alphabet^*$ are (recursively) well matched. Let us fix an input string $\word \in \alphabet^\omega$. The \emph{abstract successor} of position $i$ in $\word$, denoted $\anxt(i)$, is given as follows.
	\[
	\anxt(i) = \left\{ \begin{array}{ll}
		j & \mbox{if } \substr{i}{j+1}{\word} = c\: u\: r \\
		& \mbox{where $u$ is well matched,} \\
		& c \in \alphabet_\call, r \in \alphabet_\return\\
		\s{undefined} & \mbox{otherwise}
	\end{array} \right.
	\]
	Notice that $\anxt(i)$ is defined only if $\ith{\word} \in \alphabet_\call$, and if defined, its value is the position of its \emph{matching return symbol}. The abstract successor identifies the position of the next symbol that will be read if the 1-AJA decides to \emph{jump} on a call. Next, the \emph{local successor} of position $i$ in $\word$, denoted $\nxt(i)$, is always defined to be $i+1$. 
	
	Let us fix a 1-AJA $\auto = \autoexpanded$ and input string $\word \in \alphabet^\omega$. A \emph{run} of $\auto$ on $\word$ is a rooted, labeled tree $R = (V,E,r,L)$, where $V$ is the set of vertices, $E$ the set of edges, $r \in V$ is the root, and $L: V \to (\ctrlstates \times \natnums)$. The label of vertex indicates the state of $\auto$ and the position in $\word$ that is being read. We require that $L(r) = (\initctrlstate,0)$, i.e., the automaton is in the initial state and the $0$th symbol is being read. Let $v \in V$ be an arbitrary vertex with $L(v) = (\ctrlstate,i)$ and let $C \subseteq V$ be the set vertices that are children of $v$ in $R$. We require that the labels of $v$ and those of vertices in $C$ are consistent with $\autotrans$ --- there is a set $A \subseteq (\ctrlstates \times \set{\nxt,\anxt})$ such that (a) $A \satisfy \autotrans(\ctrlstate,\ith{\word})$; (b) for every $(\ctrlstate',d) \in A$ such that $d(i)$ is defined, there is a vertex $n \in C$ with $L(n) = (\ctrlstate',d(i))$; and (c) for every vertex $n \in C$, there is $(\ctrlstate',d) \in A$ such that $d(i)$ is defined and $L(n) = (\ctrlstate',d(i))$. A run $R = (V,E,r,L)$ is \emph{accepting} if every infinite path in $R$ satisfies the \emph{parity acceptance} condition as defined by $\parityfct$. That is, for every infinite path $\exec$ in $R$, $\s{mp}(\exec)$ is even, where
	\[
		\s{mp}(\exec) = \min \{\parityfct(\ctrlstate)\: |\: \mbox{for infinitely many } i,\ L(\ith{\exec}) = (\ctrlstate,k) \mbox{ for some }k\}.
	\]
	Finally, as always, the language recognized by $\auto$ is given by
	\[
	\lang(\auto) = \setpred{\word \in \alphabet^\omega}{\auto \text{ has an accepting run on } \word}.
	\]
	
	
	\section{Proof of \lemref{main}}
	\applabel{proof-main}
	
	Before beginning the proof of \lemref{main}, let us recall why NVPAs and 1-AJAs are closed on standard operations on languages. We present these observations with a focus on the size of the resulting automata. For a language $A \subseteq \alphabet^\omega$ and subset $\Gamma \subseteq \alphabet$, we take $\Gamma A$ to be the language $\setpred{\alphele\word}{\alphele \in \Gamma \text{ and } \word \in A}$.
	\begin{proposition}
		\proplabel{closure}
		Let $\auto_i$ be a 1-AJA (NVPA) recognizing $L_i \subseteq \alphabet^\omega$ of size $n_i$, for $i \in \set{1,2}$. Let $\Gamma \subseteq \alphabet$. Then the following automata can be constructed in polynomial time.
		\begin{enumerate}
			\item There is a 1-AJA (NVPA) recognizing $L_1 \cup L_2$ of size $O(n_1+n_2)$. 
			\item There is a 1-AJA (NVPA) recognizing $\Gamma L_1$ of size $O(n_1)$.
			\item There is a 1-AJA recognizing $(\alphabet^\omega \setminus L_1)$ of size $O(n_1)$.
		\end{enumerate}
	\end{proposition}
	\begin{proof}[Sketch]
		The automaton construction that establishes (1) chooses to either run one of $\auto_1$ or $\auto_2$ nondeterministically to recognize $L_1 \cup L_2$. (2) can be established by checking if the first symbol belongs to $\Gamma$ (and performing the necessary stack operation demanded by the first symbol in the case of NVPA) and then running the automaton for $L_1$. Finally, for (3), the 1-AJA recognizing the complement of $L_1$ has the same set of states, the parity of each state is increased by $1$, and for any state $q$ and symbol $a$, $\autotrans(q,a) = \s{dual}(\autotrans_1(q,a))$, where $\autotrans_1$ and $\autotrans$ are the transition functions of $\auto_1$ and the automaton recognizing $(\alphabet^\omega\setminus L_1)$, respectively.
	\end{proof}
	
	Our construction of $\auto_\spformula$ will proceed inductively based on the structure of $\spformula$. The type of automaton constructed will be consistent with the parity of $\alt(\spformula)$, i.e., an NVPA if $\alt(\varphi)$ is odd and a 1-AJA if $\alt(\spformula)$ is even.
		
	\mycases{$\spformula = \atomprop_\pathvar$}{%
		Let $\Gamma \subseteq \alphabet[m]$ be the set of symbols $\inpsymb$ such that $a$ is in the label set of the $\pathvar$th control state in symbol $\inpsymb$. We will construct a 1-AJA that accepts $\Gamma(\alphabet[m])^\omega$ using \propref{closure} (2).
	}
		
	\mycases{$\spformula = \lnot\spformula_1$}{%
		We will convert $\auto_{\spformula_1}$ into a 1-AJA (if needed) using \thmref{aja-vpa}, and then use \propref{closure} (3) to construct a 1-AJA that accepts the language $\overline{\lang(\auto_{\spformula_1})}$.
	}
		
	\mycases{$\spformula = \spformula_1 \orsymb \spformula_2$}{%
		Without loss of generality, assume that $\alt(\spformula_1) \geq \alt(\spformula_2)$. We convert (if needed) $\auto_{\spformula_2}$ to an automaton of the same type as $\auto_{\spformula_1}$ using \thmref{aja-vpa} and then use \propref{closure} (1), to construct an automaton recognizing $\lang(\auto_{\spformula_1}) \cup \lang(\auto_{\spformula_2})$.
	}
		
	\mycases{$\spformula = \next\spformula_1$}{%
		We use \propref{closure} (2) to construct an automaton for $\alphabet[m]\lang(\auto_{\spformula_1})$.
	}
		
	\mycases{$\spformula = \spformula_1 \until \spformula_2$}{%
	    Using \thmref{aja-vpa}, we will construct (if needed) a 1-AJA equivalent to $\auto_{\spformula_i}$ and let this be $\auto_i = (\ctrlstates_i,\ctrlstate_{\s{in}_i},\alphabet[m],\autotrans_i,\parityfct_i)$, for $i \in \set{1,2}$. The 1-AJA for $\spformula$ is given by $\auto_\spformula = (\ctrlstates_1 \disjunion \ctrlstates_2 \disjunion \set{\initctrlstate}, \initctrlstate, \alphabet[m], \autotrans, \parityfct)$ where $\autotrans(\initctrlstate,\inpsymb) = \autotrans_2(\ctrlstate_{\s{in}_2},\inpsymb) \orsymb ((\nxt,\initctrlstate) \andsymb \autotrans_1(\ctrlstate_{\s{in}_1},\inpsymb))$, $\parityfct(\initctrlstate) = 1$, and for $\ctrlstate \in \ctrlstates_i$ ($i \in \set{1,2}$), we have $\autotrans(\ctrlstate,\inpsymb) = \autotrans_i(\ctrlstate,\inpsymb)$ and $\parityfct(\ctrlstate) = \parityfct_i(\ctrlstate)$.
	}
		
	\mycases{$\spformula = \existspi\spformula_1$}{%
		Recall that our pushdown system $\PDS$ has control states $\pdsstates$, stack alphabet $\stalph$, and transition relation $\PDSrel = \PDSrelexp$. Using \thmref{aja-vpa}, we will construct (if needed) a NVPA equivalent to $\auto_{\spformula_1}$ and let this be $\auto_1 = (\ctrlstates_1,\ctrlstate_{\s{in}_1},\alphabet[m+1],\stalph_1,\stbot,\autotransVPA_1,\ctrlstates_{F_1})$. Notice that the input alphabet of $\auto_1$ is $\alphabet[m+1]$, where the $m+1$st component is an encoding of the path assigned to variable $\pathvar$. The automaton for $\spformula$ will essentially guess the encoding of a path that is consistent with the transitions of $\PDS$, and check if assigning the guessed path to variable $\pathvar$ satisfies $\spformula_1$ by running the automaton $\auto_1$. The additional requirement we have is that the guessed path start at the \emph{same configuration} as the current configuration of the path assigned to variable $\pathvarsele_m$. In order to be able to guess a path, $\auto_\spformula$ will keep track of $\PDS$'s control state in its control state, and use its stack to track $\PDS$'s stack operations along the guessed path. Before defining $\auto_\spformula$ formally we need to introduce some notation that will be convenient. An element $\inpsymb \in \alphabet[m]$ is of the form $(\s{o},(\pdsstate_1,\pdsstate_2,\ldots \pdsstate_m),(\stele_1,\stele_2,\ldots\stele_m))$. We use $\s{op}(\inpsymb)$ to denote $\s{o}$, $\proj[i]{\state(\inpsymb)}$ to denote $\pdsstate_i$, and $\proj[i]{\stack(\inpsymb)}$ to denote $\stele_i$, for $i \in \set{1,2,\ldots m}$. By convention we take $\proj[0]{\state(\inpsymb)} = \initpdsstate$, and $\proj[0]{\stack(\inpsymb)}$ to be \emph{undefined}. Finally, for $\pdsstate \in \pdsstates$ and $\stele \in \stalph$, $\inpsymb + (\pdsstate,\stele)$ is the symbol in alphabet $\Sigma[m+1]$ given by $(\s{o}, (\pdsstate_1,\ldots \pdsstate_m, \pdsstate), (\stele_1,\ldots \stele_m, \stele))$.
			
		The NVPA $\auto_\spformula = ((\ctrlstates_1 \times \pdsstates) \disjunion \set{\initctrlstate}, \initctrlstate, \alphabet[m], (\stalph_1 \times \stalph), \stbot, \autotransVPA, (\ctrlstates_{F_1} \times \pdsstates))$; notice that the bottom of stack symbol ($\stbot$) is same as the one in $\auto_1$. The transition relation $\autotransVPA = \autotransexpandedVPA$ is defined as follows.
		\begin{itemize}
			\item When $\s{op}(\inpsymb) = \call$,
			\[
			\begin{array}{rl}
				\autotransVPA_\call = & \{(\initctrlstate,\inpsymb,(\ctrlstate,\pdsstate),(b,\stele))\: |\ (\ctrlstate_{\s{in}_1},\inpsymb + (\proj{\state(\inpsymb)},\stele), \ctrlstate,b) \in \autotransVPA_1 \mbox{ and }\\
				& \hspace*{3.3cm} (\proj{\state(\inpsymb)},(\pdsstate,\stele)) \in \PDSrel_\call\}\ \cup\\
				& \{((\ctrlstate,\pdsstate),\inpsymb,(\ctrlstate',\pdsstate'),(b,\stele))\: |\ (\ctrlstate,\inpsymb + (\pdsstate,\stele), \ctrlstate',b) \in \autotrans_1,\ (\pdsstate,(\pdsstate',\stele)) \in \PDSrel_\call\}.
			\end{array}
			\]
			\item When $\s{op}(\inpsymb) = \internal$,
			\[
			\begin{array}{rl}
				\autotransVPA_\internal = & \{(\initctrlstate,\inpsymb,(\ctrlstate,\pdsstate))\: |\: (\ctrlstate_{\s{in}_1},\inpsymb + (\proj{\state(\inpsymb)},\emptyword), \ctrlstate) \in \autotransVPA_1,\ (\proj{\state(\inpsymb)},\pdsstate) \in \PDSrel_\internal\}\\
				& \cup\: \{((\ctrlstate,\pdsstate),\inpsymb,(\ctrlstate',\pdsstate'))\: |\: (\ctrlstate,\inpsymb + (\pdsstate,\emptyword), \ctrlstate') \in \autotrans_1,\mbox{ and } (\pdsstate,\pdsstate') \in \PDSrel_\internal\}.
			\end{array}
			\]
			\item When $\s{op}(\inpsymb) = \return$,
			\[
			\begin{array}{rl}
				\autotransVPA_\return = & \{(\initctrlstate,\inpsymb,\stbot,(\ctrlstate,\pdsstate))\: |\ (\ctrlstate_{\s{in}_1},\inpsymb + (\proj{\state(\inpsymb)},\proj{\stack(\inpsymb)}), \stbot, \ctrlstate) \in \autotransVPA_1 \mbox{ and }\\
				& \hspace*{2.8cm} ((\proj{\state(\inpsymb)},\proj{\stack(\inpsymb)}),\pdsstate) \in \PDSrel_\return\}\ \cup\\
				& \{((\ctrlstate,\pdsstate),\inpsymb,\stbot,(\ctrlstate',\pdsstate'))\: |\ (\ctrlstate,\inpsymb + (\pdsstate,\proj{\stack(\inpsymb)}), \stbot, \ctrlstate') \in \autotransVPA_1, \mbox{ and}\\
				& \hspace*{3.4cm} ((\pdsstate,\proj{\stack(\inpsymb)}),\pdsstate') \in \PDSrel_\return\}\ \cup\\
				& \{((\ctrlstate,\pdsstate),\inpsymb,(b,\stele),(\ctrlstate',\pdsstate'))\: |\ (\ctrlstate,\inpsymb + (\pdsstate,\stele), b, \ctrlstate') \in \autotransVPA_1,\ ((\pdsstate,\stele),\pdsstate') \in \PDSrel_\return\}
			\end{array}
			\]
		\end{itemize}
		The requirement that the guessed path for $\pathvar$ start in the same configuration as the current configuration of the path assigned to $\pathvarsele_m$, introduces a few points in the definition of $\autotransVPA$ that are worth highlighting. Transitions from $\initctrlstate$, whether they be $\call,\internal$, or $\return$, pick a step in $\PDS$ that starts from the same state as the one in the path assigned to $\pathvarsele_m$. Stack symbols pushed by $\PDS$ along the guessed path are pushed by $\auto_\spformula$ onto its stack (see $\autotransVPA_\call$). If the stack of $\auto_\varphi$ is $\stbot$ at a $\return$-transition, that means on the guessed path the symbol popped must be from what was on the stack at the start of the path. Since that matches with the configuration on the path mapped to $\pathvarsele_m$, this symbol must be the same as what is popped for $\pathvarsele_m$. This is reflected in $\autotransVPA_\return$ for the cases with $\stbot$.
			
		An important observation that we will exploit is that if $\spformula = \existspi\spformula_1$ is a sentence, then the following stronger correctness guarantee holds: for any $\stprof \in \set{\call,\internal,\return}^\omega$, $\stprof \in \lang(\auto_{\varphi})$ if any only if $\stprof$ is a stack access pattern and $\PDS, [\spsymbol \mapsto \stprof], \spsymbol \satisfy \spformula$. The language of $\auto_\spformula$ in this case consists of exactly the set of path environments satisfying $\spformula$. This stronger statement follows from the construction of $\auto_\spformula$.
		}
		
	To complete the proof, we will bound the size of $\auto_\spformula$ by induction on $\alt(\spformula)$. Recall that $n$ is a bound on the size of $\PDS$ and $\spformula$. 
	    
	\mycases[Base]{Case}{%
	    Consider $\spformula$ such that $\alt(\spformula) = 0$. Based on the definition of $\alt(\cdot)$ (\defref{alt-depth}), this means that $\spformula$ is built from propositions, $\lnot$, $\orsymb$, $\next$, and $\until$; in particular there are no path quantifiers in $\spformula$ in this case. Observe that the construction for $\atomprop_\pathvar$ is a constant sized automaton. Also, the constructions for $\lnot$, $\orsymb$, $\next$, and $\until$ add at most a constant factor to the size of the automaton. Given these observations, size of $\auto_\spformula$ in this case is bounded by $O(n)$, which is bounded by $g_{O(1)}(0,n)$. Thus the base case holds.
	}
	    
	\mycases[Induction]{Step}{%
	    We break the induction step into two cases based on the parity of $\alt(\spformula)$. When $\alt(\spformula) = 2k$ (for some $k \geq 1$) then $\spformula$ is built using sub-formulas of 
	    {(formula) complexity} at most $2k$ using Boolean operators ($\lnot,\orsymb$), $\next$, and $\until$. Let us assume that we first construct 1-AJAs for each of the subformulas with 
	    {(formula) complexity} $< 2k$. This results in at most a quadratic blowup (\thmref{aja-vpa}), and so the size of the automata for each such subformula is at most $(g_c(k,n))^2$ for some $c$. The constructions for $\lnot,\orsymb,\next$ and $\until$ produce an automaton that is at most a constant factor of the sum of the sizes of the component automata. Thus, the size of $\auto_\spformula$ is at most $dn(g_c(k,n))^2 \leq g_{c'}(k,n)$ for some $c'$, establishing the claim in this case. Now let us consider the case when $\alt(\spformula) = 2k+1$ for some $k \geq 0$. In this case $\spformula$ is built using $\orsymb$, $\next$ and $\exists$ to combine subformulas of 
	    {(formula) complexity} $\leq 2k+1$. Again, we can convert automata for each of the sub-formulas of 
	    {(formula) complexity} $< 2k+1$ into NVPAs for an exponential cost (\thmref{aja-vpa}). Thus, we can assume that the size of all of these automata is bounded by $g_c(k+1,n)$ for some $c$. Based on \propref{closure}, we can see that disjunction and $\next$ produce automata that grow by at most a constant factor. Existential quantification are the only operators to have a non-trivial blow-up. Based on the construction outlined in this proof, if $\varphi$ has an NVPA of size $\ell$ then the automaton of $\exists\pathvar.\: \varphi$ has size $O(n\ell)$ as $n$ bounds the size of $\PDS$. Since there are at most $n$ quantifiers, we can bound the size of $\auto_\spformula$ by $dn^n(g_c(k+1,n)) \leq 2^{d'n\log n + g_c(k,n)} \leq g_{c'}(k+1,n)$; the last step is because $g_c(k,n) \geq cn\log n$. This establishes the induction step.
	}
	
	\section{Proof of \thmref{lower-bound}}
	\applabel{lower-bound-proof}
	
		We will show that every language $L \in \aspace{h_c(k-1,n)}$ can be reduced to the model checking problem of {\rmodltl} sentence with formula complexity $2k-1$. Since $\aspace{f(n)} = \dtime{2^{O(f(n))}}$ for $f(n) \geq \log n$, the theorem will follow.
	    
	    Consider an arbitrary $h_c(k-1,n)$-space bounded alternating Turing machine (ATM) $M$. Since $h_c(k-1,n) \geq n$, we may assume that $M$ is a 1-tape machine. Let $M = (\ctrlstates_\exists, \ctrlstates_\forall, \alphabet, \stalph, \sqcup, \initctrlstate, \autotrans, \ctrlstate_a)$, where $\ctrlstates_\exists$ is the set of \emph{existential} control states, $\ctrlstates_\forall$ is the set of \emph{universal} control states, $\alphabet$ is the \emph{input alphabet}; $\stalph \supseteq \alphabet$ is the \emph{tape alphabet}, $\sqcup \in \stalph \setminus \alphabet$ is the \emph{blank symbol}, $\initctrlstate \in \ctrlstates_\exists \cup \ctrlstates_\forall$ and $\ctrlstate_a \in \ctrlstates_\exists$ are the \emph{initial} and \emph{accepting} states, respectively, and $\autotrans$ is the transition function of the Turing machine. We use $\ctrlstates = \ctrlstates_\exists \cup \ctrlstates_\forall$ to denote the set of all states of $M$. We assume that $\ctrlstate_a$ is a halting state (no transitions enabled) and so $\autotrans: (\ctrlstates \setminus \set{\ctrlstate_a}) \times \stalph \to 2^{(\ctrlstates \times \stalph \times \set{\rght,\lft})}$, where given a state and current symbol being read, the transition function identifies choices for the next state, the symbol to be written, and the direction in which to move the tape head ($\rght$ for right, and $\lft$ for left). We assume, without loss of generality, that for each pair $(\ctrlstate,b) \in \ctrlstates \times \stalph$ that $|\autotrans(\ctrlstate,b)| \in \set{0,2}$, i.e., there are either no or two choices at each step. Also, we assume that these choices are ordered in some fashion so we will often speak of ``choice $i$'' for $i \in \set{1,2}$.
	    
	    A \emph{configuration} $\config$ of $M$ is a string in $\stalph^*(\ctrlstates\times\stalph)\stalph^*$, where $\config = u(\ctrlstate,b)v$ with $u,v \in \stalph^*$, $b \in \stalph$ and $\ctrlstate \in \ctrlstates$, denotes that the tape of $M$ is the string $ubv$, the control state is $\ctrlstate$, and the head is reading the cell containing $b$. Since $M$ is $h_c(k-1,n)$-space bounded, we can assume any configuration $\config$ of $M$ is a string of length \emph{exactly} $h_c(k-1,n)$. The initial configuration of $M$ on input $bw$ of length $n$ is $(\initctrlstate,b) w\sqcup^{h_c(k-1,n)-n}$. A configuration $\config = u(\ctrlstate,b)v$ is an \emph{existential} configuration if $\ctrlstate \in \ctrlstates_\exists$, a \emph{universal} configuration if $\ctrlstate \in \ctrlstates_\forall$, and an \emph{accepting} configuration if $\ctrlstate = \ctrlstate_a$. For a pair of configurations $\config$ and $\config'$, we say $\config \tmmv \config'$ if $M$'s configuration is $\config'$ if it takes one step according to choice $i \in \set{1,2}$ from configuration $\config$. A \emph{run} of $M$ on input $w$ is a \emph{finite}, rooted binary tree $T=(V,E,r,\ell)$, where $V$ is the set of vertices, $E$ is the set of edges oriented away from the root, $r \in V$ is the root, and $\ell$ is a function that maps each vertex to a configuration of $M$. In addition, $\ell$ is required to satisfy the following conditions. The root $r$ is labeled by the initial configuration. For any internal vertex $v \in V$, if $\ell(v)$ is an existential configuration then $v$ has one child $c$ such that $\ell(v) \tmmv \ell(c)$ for some $i \in \set{1,2}$, and if $\ell(v)$ is a universal configuration, then $v$ has two children $c_1,c_2$ such that $\ell(v) \tmmv \ell(c_i)$ for $i \in \set{1,2}$. Finally, a run $T$ is \emph{accepting} if every leaf of $T$ is labeled by an accepting configuration. An input $w$ is accepted by $M$, if $M$ has an accepting run on $w$.
	    
	    We will construct a reduction from $\lang(M)$ (which by definition is in $\aspace{h_c(k-1,n}$) to the model checking problem for {\rmodltl}. That is, given input $w$, we will construct a pushdown system $\PDS_w$ and {\rmodltl} sentence $\ctlformula_w$ such that $\PDS_w \satisfy \ctlformula_w$ if and only if $w \in \lang(M)$. The idea behind the reduction is to construct a pushdown system $\PDS_w$ such that labels of paths starting from the initial configuration in $\transPDS[\PDS_w]$ encode possible computations of $M$ on $w$. And $\ctlformula_w$ is constructed to check if a path encodes a valid accepting run of $M$ on $w$. To formalize this intuition, we need to first identify a way to encode runs of $M$, which are binary trees, as a string of labels.
	    
	    \paragraph{Encoding ATM Runs.}
	    Recall that a run of $M$ is a binary tree and we need to find a way to encode the tree as string. One way to accomplish this faithfully, is to have the encoding record the stack operations during a depth first search (DFS) traversal of the tree. For example, if we have a tree $T$ with root $r$, with tree $T_1$ as the left child and $T_2$ as the right child, then during DFS, the algorithm will first push $r$ on the stack, perform DFS traversal on $T_1$ (recursively), pop $r$ from the stack, push $r$ back onto the stack, DFS traverse $T_2$ (recursively), and finally pop $r$. When encoding runs, what we need to push/pop is not the node $r$, but rather its label. Notice that for a sequence of stack operations to conform to the DFS traversal of a tree, it is necessary for the symbols being pushed and popped be the label of the \emph{same} node --- for example, in the example tree before, after traversing $T_1$, we need the same node $r$ to be popped and pushed. Since the symbols when popping are in reverse order of when they are pushed, for long labels (as in the case of configurations) this check is challenging. To overcome this, we push/pop the label and its reverse at the same time. This ensures that if we want to check if a string pushed is the same as a string that was just popped, then we can check for string \emph{equality} as opposed to one being the reverse of another, and this check is easier to do using formulas in {\rmodltl}. 
	    
	    We formalize the above discussion to give a precise definition of the encoding of a binary tree as a string. We will abuse notation and overload $\enc(\cdot)$ to refer to multiple functions --- the context will disambiguate which $\enc(\cdot)$ we are referring to. Let $\Lambda = \stalph \cup (\ctrlstates \times \stalph)$, the alphabet used to encode configurations of $M$. For $i \in \set{1,2}$, let $\cpy{i}{\Lambda} = \setpred{\cpy{i}{\stele}}{\stele \in \Lambda}$ be a ``copy'' of the alphabet $\Lambda$. For a string $\config \in \Lambda^*$ of length $m$, we define
	    \[
	    \begin{array}{rl}
	    \enc(\s{push},\config) = & (\call, \cpy{1}{\ith[0]{\config}},\cpy{2}{\ith[m-1]{\config}})(\call, \cpy{1}{\ith[1]{\config}},\cpy{2}{\ith[m-2]{\config}})\cdots\\ 
	     & (\call, \cpy{1}{\ith{\config}}, \cpy{2}{\ith[m-i-1]{\config}}) \cdots (\call,\cpy{1}{\ith[m-1]{\config}},\cpy{2}{\ith[0]{\config}})\\
	    \enc(\s{pop},\config) = & (\return, \cpy{1}{\ith[m-1]{\config}},\cpy{2}{\ith[0]{\config}})(\return, \cpy{1}{\ith[m-2]{\config}},\cpy{2}{\ith[1]{\config}})\cdots\\
	     & (\return, \cpy{1}{\ith{\config}}, \cpy{2}{\ith[m-i-1]{\config}})\cdots (\return,\cpy{1}{\ith[0]{\config}},\cpy{2}{\ith[m-1]{\config}})
	    \end{array}
	    \]
	    Essentially, $\enc(\cdot)$ of a string encodes both the string and its reverse, has a tag that indicates whether the string is being pushed or popped, and if it is popped, the order of symbols is reversed. Consider a rooted, labeled binary tree $T = (V, E, r, \ell)$. Its encoding is inductively defined as follows. If $V = \set{r}$ and $E = \emptyset$ (i.e., $T$ is a tree with only one vertex) then $\enc(T) = \enc(\s{push},\ell(r))\enc(\s{pop},\ell(r))$. If $r$ has only one child which is the subtree $T_1$, then $\enc(T) = \enc(\s{push},\ell(r))\enc(T_1)\enc(\s{pop},\ell(r))$, where $\enc(T_1)$ is given recursively. Finally, if $r$ has $T_1$ and $T_2$ as left and right subtrees, respectively, then $\enc(T) = \enc(\s{push},\ell(r))\enc(T_1)\enc(\s{pop},\ell(r))\enc(\s{push},\ell(r))\enc(T_2)\enc(\s{pop},\ell(r))$. Before moving on, let us highlight a subtle aspect of our encoding. In the case of a tree where $r$ has two sub-trees $T_1$ and $T_2$, we ``pop'' $\ell(r)$ and ``push'' $\ell(r)$ between the traversals of $T_1$ and $T_2$. This may seem unnecessary on first reading. Notice that for $T$ to be a valid run, the label of the root of $T_2$ must be the result of taking one step from $\ell(r)$. Such checks will be encoded in our sentence, and for that to be possible, we need successive ATM configurations to consecutive in the string encoding.
	    
	    \paragraph{The Pushdown System.}
	    Labels of paths of our constructed pushdown system will encode possible runs of $M$ on the input $w$. At each step the pushdown system guesses the next symbol of the possible run by moving to a control state whose label corresponds to this symbol. The stack is used to ensure that when the ``pop'' symbols in the encoding are encountered they match the symbols that were ``pushed'' earlier in the guess. We can define this precisely as follows. Recall that for $\Lambda = \stalph \cup (\ctrlstates \times \stalph)$, $\cpy{i}{\Lambda}$ ($i \in \set{1,2}$) refers to the ``$i$th copy'' of $\Lambda$. Let us fix the set of atomic propositions $\atomprops = \set{\call,\return,\s{end}} \cup \cpy{1}{\Lambda} \cup \cpy{2}{\Lambda}$ and let $\pdsstates = \set{\call,\return} \times \cpy{1}{\Lambda} \times \cpy{2}{\Lambda}$. Let $\PDS = (\pdsstates \cup \set{\initpdsstate,\pdsstate_e}, (\cpy{1}{\Lambda} \times \cpy{2}{\Lambda}) \cup \set{\stbot}, \initpdsstate, \PDSrel, \labelfct)$, where $\labelfct(\initpdsstate) = \emptyset$, $\labelfct(\pdsstate_e) = \set{\s{end}}$, and $\labelfct((\s{o},\cpy{1}{a},\cpy{2}{b})) = \set{\s{o},\cpy{1}{a},\cpy{2}{b}}$. The transition relation $\PDSrel = \PDSrelexp$ is given as follows: $\PDSrel_\internal = \set{(\pdsstate_e,\pdsstate_e)}$ and
	    \[
	    \begin{array}{rl}
	    \PDSrel_\call = & \setpred{(\initpdsstate, (\pdsstate,\stbot))}{\pdsstate \in \pdsstates} \union\\
	     & \setpred{((\call,\cpy{1}{a},\cpy{2}{b}), (\pdsstate, (\cpy{1}{a},\cpy{2}{b})))}{\pdsstate \in \pdsstates \mbox{ and } a,b \in \Lambda}\\
	    \PDSrel_\return = & \setpred{(((\return,\cpy{1}{a},\cpy{2}{b}), (\cpy{1}{a},\cpy{2}{b})),\pdsstate)}{\pdsstate \in \pdsstates \mbox{ and } a,b \in \Lambda} \union\\
	     & \setpred{((\pdsstate,\stbot),\pdsstate_e)}{\pdsstate \in \pdsstates}
	    \end{array}
	    \]
	    Paths of $\transPDS$ may not correspond to actual computations of $M$ on $w$ since $\PDS$ doesn't check for many properties that need to hold for such a string. On the other hand, correct runs of $M$ on $w$ do correspond to paths of $\PDS$ that end in $(\s{end})^\omega$. Our final pushdown system will be a slight modification of $\PDS$, in two ways. First we will add some additional book-keeping to the states and stack symbols to ensure that whenever a universal configuration is guessed, the computation has transitions corresponding to both choices. Second, we will need to modify the system to account for the specification, as we shall see towards the end of this proof.
	    
	    
	    As mentioned before, for the labels of a path of $\PDS$ to correspond to the encoding of an accepting computation of $M$ on $w$, we need to ensure that the labels satisfy a few properties. These will be encoded in our {\rmodltl} sentence. However, instead of encoding these conditions in {\rmodltl}, we will find it convenient to first write them in {\qptl}, a logic introduced in~\cite{qptl-sistla}. We begin by introducing this logic, showing how the properties of accepting runs can be encoded, and then describing a way to translate them back to {\rmodltl}.
	    
	    \paragraph{{\qptl}.}
	    Quantified propositional temporal logic (QPTL)~\cite{qptl-sistla} extends LTL with quantification over propositions. Fixing a set of atomic propositions $\atomprops$, formulas in the logic are given by the following BNF grammar; in what follows, $\atomprop$ is an element of $\atomprops$.
	    \[
		\begin{array}{ccccccccccccccc}
			\varphi & \coleq & \atomprop & \text{ $\vert$ } & \lnot\varphi & \text{ $\vert$ } & \varphi \orsymb \varphi & \text{ $\vert$ } & \next \varphi & \text{ $\vert$ } & \diamondsymb\varphi & \text{ $\vert$ } & \exists \atomprop .\: \varphi 
		\end{array}
	    \]
	    Models of QPTL are the same as those for LTL, namely elements of $(2^{\atomprops})^\omega$, and the semantics of most of the constructs is similar. For $\word \in (2^{\atomprops})^\omega$, $\atomprop$ holds if $\atomprop \in \ith[0]{\word}$; $\lnot\varphi$ holds if $\varphi$ does not hold; $\varphi_1 \orsymb \varphi_2$ holds if either $\varphi_1$ or $\varphi_2$ hold; $\next\varphi$ holds if $\varphi$ holds on the suffix $\suffix[1]{\word}$; and $\diamondsymb \varphi$ holds if $\varphi$ holds in some suffix $\suffix{\word}$. The only new operator is $\exists \atomprop.\: \varphi$ which holds in $w$, if $\varphi$ holds in some word $w'$ which agrees with $w$ in the evaluation of all propositions except possibly $\atomprop$. Recall that $\diamondsymb\varphi$ is equivalent to $\tru\until\varphi$, $\squaresymb\varphi$ is a short hand for $\lnot\diamondsymb(\lnot\varphi)$, and $\forall\atomprop.\:\varphi$ is $\lnot\exists\atomprop.\: (\lnot \varphi)$. We can extend $\alt(\cdot)$ to {\qptl} formulas, with the same definition; for this definition $\lnot$ will behave like negation for {\stprofadj} formulas, rather than negation for {\modltl}-sentences. Finally, it has been shown that every {\qptl} formula is equivalent to one in prenex normal form, where all the quantifiers have been pulled to the front of the formula.
	    
	    Our {\qptl} formula $\varphi_w$ that describes when a word encodes an accepting run of $M$ on $w$, will rely on formulas constructed in~\cite{qptl-sistla}. The first is formula $\varphi_{c,k,n}(p_1,p_2)$ (Lemma 4.4. in~\cite{qptl-sistla}) of size $O(k+n)$ such that $\word \satisfy \varphi_{c,k,n}(p_1,p_2)$ if and only if propositions $p_1$ and $p_2$ are true exactly once in $\word$, $p_2$ is true after $p_1$, and they are separated by exactly $h_c(k,n)$ positions. Further more $\alt(\varphi_{c,k,n}) = 2k-1$, and can be constructed $O(\log n)$ space. We will introduce the other formulas as we describe the conditions needed.
	    
	    \paragraph{The {\qptl} formula $\varphi_w$.}
	    We now describe $\varphi_w$ with the property that a path of $\transPDS$ satisfies $\varphi_w$ if and only if $M$ accepts $w$; here a path $\config_0\config_1\cdots$ of $\transPDS$ satisfies $\varphi_w$, if $\labelfct(\config_0)\labelfct(\config_1)\cdots \satisfy \varphi_w$. Observe that the construction of $\PDS$ ensures that symbols ``popped'' are the same as the symbols ``pushed'' and that every universal configuration has two successors correspond to transitions corresponding to each choice.. Therefore, the remaining conditions that need to be checked for a path to be the encoding of an accepting computation of $M$ on $w$ are as follows.
	    \begin{enumerate}
	        \item\label{it1} Every configuration has length exactly $h_c(k-1,n)$.
	        \item\label{it2} All symbols encoding a single configuration have the same tag, i.e., either all $\call$ or all $\return$.
	        \item\label{it3} The first configuration is the initial configuration of $M$ on $w$.
	        \item\label{it4} Successive ``pushed'' configurations correspond to a single step of $M$ consistent with its transition function.
	        \item\label{it5} ``Leaves'' of the run are accepting configurations. In other words, if a push configuration is immediately followed by a pop configuration (they will be identical thanks to $\PDS$), then the control state must be $\ctrlstate_a$.
	        \item\label{it6} If a pop configuration  is immediately followed by a push configuration (i.e., the DFS traversal of the run is exploring the right child) then they must be the same configuration.
	        \item\label{it7} All the configurations are popped at the end.
	    \end{enumerate}
	    If $\varphi$ denotes the conjunction of all the above conditions written in {\qptl}, then our desired formula $\varphi_w$ is $\next\varphi$ since 
	    the initial state $\initpdsstate$ of $\PDS$ is not part of guessing the encoding of the computation.
	    
	    We describe how to encode each of these conditions in {\qptl}, often relying on formulas from~\cite{qptl-sistla}.
	    \begin{enumerate}
	        \item To state property (\ref{it1}), we will have a proposition $\sep$ (which will be existentially quantified) that marks the beginning of each configuration. The fact that $\sep$ marks the beginning of each configuration will be ensured by the other properties we write down. We will require that $\sep$ is true exactly $h_c(k-1,n)$ positions apart using the formula $\varphi_{c,k-1,n}(\cdot,\cdot)$ introduced before. This is given by Equation (9) in~\cite{qptl-sistla} where $r$ replaced by $\sep$.
	        \item Since $\sep$ marks the beginning of each configuration, saying that all symbols in a configuration have the same tag, is equivalent to saying that if the tag changes then $\sep$ must be true when the tag changes. In other words, property (\ref{it2}) can be written as
	        \[
	        \squaresymb ((\call \andsymb \next \return) \impsymb \next\sep) \andsymb \squaresymb ((\return \andsymb \next \call) \impsymb \next\sep)
	        \]
	        \item Equation (10) in~\cite{qptl-sistla} states the property (\ref{it3}). It needs to be slightly modified to also include the condition that the propositions from $\cpy{2}{\Lambda}$ encode the reverse of the initial configuration of $M$.
	        \item Equation (11) in~\cite{qptl-sistla} states the property (\ref{it4}). It needs to be slightly to modified to also ensure that the reverse encodings using $\cpy{2}{\Lambda}$ are consistent with the transitions of $M$, and this requirement is only imposed for successive configurations that have the $\call$ tag.
	        \item Recall that corresponding symbols in successive configurations are $h_c(k-1,n)$ apart. Property (\ref{it5}) can be written to say that if there are two positions $p$ and $q$ that are $h_c(k-1,n)$ apart that have opposite tags, and if in addition $p$ corresponds to a position that is scanned by the tape head, then the control state at $p$ must be the accepting state $\ctrlstate_a$.
	        \begin{multline*}
	        \forall p_1\forall p_2.\ (\varphi_{c,k-1,n}(p_1,p_2) \andsymb \diamondsymb (p_1 \andsymb \call \andsymb \bigvee_{a \in \ctrlstates \times \stalph}\cpy{1}{a}) \andsymb \diamondsymb(p_2 \andsymb \return)) \impsymb\\
	        \diamondsymb (p_1 \andsymb \bigvee_{a \in \set{\ctrlstate_a} \times \stalph} \cpy{1}{a})
	        \end{multline*}
	        \item Using the observations in property (\ref{it5}), property (\ref{it6}) can be written as
	        \begin{multline*}
	        \forall p_1\forall p_2.\ ((\varphi_{c,k-1,n}(p_1,p_2) \andsymb \diamondsymb (p_1 \andsymb \return) \andsymb \diamondsymb (p_2 \andsymb \call)) \impsymb\\
	        \bigvee_{a,b \in \Lambda} (\diamondsymb (p_1 \andsymb \cpy{1}{a} \andsymb \cpy{2}{b}) \andsymb \diamondsymb (p_2 \andsymb \cpy{1}{b} \andsymb \cpy{2}{a}))
	        \end{multline*}
	        \item This property can be ensured by requiring that the path end in state $\pdsstate_2$. We write this as $\diamondsymb\s{end}$.
	    \end{enumerate}
	    Since formulas in {\qptl} can be written in prenex normal form, we can pull all the universal quantifiers in the various formulas listed above to get a formula $\varphi_w$ of the form $\next\:\exists\sep\forall p_1\forall p_2.\ \varphi$ where $\varphi$ is a Boolean combination of $\lnot\varphi_{c,k-1,n}(\cdot,\cdot)$ and simple formulas with temporal operators of 
	    {(formula) complexity} $0$. Thus, $\alt(\varphi_w) = 2k-1$.

    \paragraph{Converting to {\rmodltl}.}
    Our {\qptl} formula $\varphi_w$ is of the form $\next\exists\sep\varphi'$. We will describe how to construct an ``equivalent'' formula in {\rmodltl}; this will require modifying our pushdown system $\PDS$ as well to obtain our final pushdown system $\PDS_w$. Construct a {\stprofadj} sentence $\spformula'$ as follows. Let $\atomprop$ be a new proposition \emph{not appearing} in $\varphi_w$. For every \emph{quantified} proposition $p$ in $\varphi_w$, replace every bound occurrence of $p$ by $\atomprop_{\pathvar_p}$ and replace the quantification $\exists p$ with $\exists \pathvar_p$. Let $\pathvar_*$ be a fresh path variable. Replace every free proposition $p$ in $\varphi_w$ by $p_{\pathvar_*}$. Let $\spformula$ be the resulting {\stprofadj} formula. Finally let $\ctlformula_w = E\exists\pathvar_*.\ \spformula$.
    
    To finish the reduction, we need to modify $\PDS$ to account for the new proposition $\atomprop$ introduced in $\ctlformula_w$. Our final pushdown system $\PDS_w$ is constructed as follows. Recall that $\pdsstates = (\set{\call,\return} \times \cpy{1}{\Lambda} \times \cpy{2}{\Lambda})$ and the states of $\PDS$ is $\pdsstates \cup \set{\initpdsstate,\pdsstate_e}$. Let $\cpy{1}{\pdsstates}$ and $\cpy{2}{\pdsstates}$ be two copies of $\pdsstates$. The states of $\PDS_w$ will be $\cpy{1}{\pdsstates} \cup \cpy{2}{\pdsstates} \cup \set{\initpdsstate,\pdsstate_e}$. The two copies of state $\pdsstate \in \pdsstates$ will have the same transitions into and out of it and have the same labels for all the old propositions. The only difference will be that $\atomprop \in \labelfct(\cpy{1}{\pdsstate})$ while $a \not\in \labelfct(\cpy{2}{\pdsstate})$. 
    
    It is easy to make the following observations: $\alt(\ctlformula_w) = \alt(\varphi_w)$, and a path of $\transPDS$ satisfies $\varphi_w$ if and only if $\PDS_w \satisfy \ctlformula_w$. This completes the proof of the hardness result.
\else 
\end{document}
\fi
\end{document}